\documentclass[twocolumn]{aastex701}
\usepackage{lineno}
\linenumbers
\usepackage{hhline}
\usepackage{amsmath,amssymb}
\usepackage{bm}
\usepackage{subcaption}
\usepackage{graphicx}
\usepackage{lipsum}
\usepackage{gensymb}
\usepackage{comment}

\newcommand{\RZ}{R_{Z^2}}
\newcommand{\Delsig}{\Delta\sigma_c}

\begin{document}

\title{Solar-Cycle Modulation of Turbulence in the Young Solar Wind and the Influence of Source Regions}

\author[orcid=0000-0003-4736-4947]{Yanwen Wang}
\altaffiliation{These authors contributed equally to this work.}
\affiliation{Department of Physics, University of Maryland, College Park, MD, USA}
\email[show]{ywwang@udel.edu}

\author[orcid=0000-0002-7174-6948]{Rohit Chhiber}
\altaffiliation{These authors contributed equally to this work.}
\affiliation{Department of Physics and Astronomy, University of Delaware, Newark, DE, USA}
\affiliation{Heliophysics Science Division, NASA Goddard Space Flight Center, Greenbelt, MD, USA}
\email[show]{rohit.chhiber@nasa.gov}

\author[orcid=0000-0002-0209-152X]{Arcadi V. Usmanov}
\affiliation{Department of Physics and Astronomy, University of Delaware, Newark, DE, USA}
\affiliation{Heliophysics Science Division, NASA Goddard Space Flight Center, Greenbelt, MD, USA}
\email{avusmanov@gmail.com}

\author[orcid=0000-0002-5435-3544]{Marc Swisdak}
\affiliation{Institute for Research in Electronics and Applied Physics, University of Maryland, College Park, MD, USA}
\email{swisdak@umd.edu}  

\author[orcid=0000-0001-7224-6024]{William H. Matthaeus}
\affiliation{Department of Physics and Astronomy, University of Delaware, Newark, DE, USA}
\email{whm@udel.edu}


\begin{abstract}

\added{Turbulence is central to the dynamics and evolution of the solar wind during its expansion from the corona into the heliosphere. However, the solar-cycle modulation of turbulence properties and their relationship to solar source structure remain incompletely understood. Here, we employ seven years of Parker Solar Probe (PSP) data, together with a global solar-wind model, photospheric magnetograms, and extreme-ultraviolet maps of the low corona, to examine both aspects of this problem. Analysis of PSP observations between 0.15 and 0.5 au reveals pronounced trends with solar activity in fluctuation energy, cross helicity, turbulent energy-transfer rate, and other relevant properties over approximately half an activity cycle. We further investigate the connection between these properties and inferred solar source regions during three PSP encounters, E6, E14, and E22, spanning different levels of solar activity. These case studies suggest that, at least during some epochs, intervals with relatively high turbulent energy and Alfv\'enicity tend to originate from coronal-hole sources with unipolar magnetic topology, whereas intervals with relatively low values of these quantities tend to be associated with multipolar topology and active-region source environments. These results demonstrate the potentially significant influence of solar-cycle--dependent source structure on turbulence variability and the associated plasma heating in the young solar wind. The absence of a clear source-region imprint during other epochs, however, motivates broader statistical analyses and more detailed case studies to establish the conditions under which heliospheric turbulence retains information about its solar origins.}

\end{abstract}



\section{Introduction} \label{sec:intro}
The solar wind is a magnetized, weakly-collisional plasma outflow from the Sun, that contains ubiquitous, broadband fluctuations in velocity, magnetic field, and density. 
These fluctuations are not a passive component of the flow: nonlinear interactions transfer energy from large to small scales, where turbulent energy can be converted to heat and contribute to the non-adiabatic thermal evolution of the solar wind \citep{matthaeus2011SSR, 
bruno2013LRSP,
verscharen2019multi}.
Analyses based on in-situ observations of the inertial-range cascade rate 
\citep[e.g.,][]{macbride2008ApJ} and turbulence-transport modeling \citep{matthaeus1999turbulence} indicate that the turbulent cascade can provide the heating required to maintain observed solar-wind temperature profiles. 
Turbulent fluctuations can also enhance pressure gradients that directly accelerate the wind to observed fast speeds \citep{usmanov2000global,Rivera2024Sci,Usmanov2025ApJ}, and induce significant scattering of solar energetic particles (SEPs) and galactic cosmic rays \citep[][]{Engelbrecht2022SSR}. \added{They can also influence the dynamics of interplanetary coronal mass ejections \citep{Telloni2026ApJ} and contribute to space-weather impacts  \citep{Damicis2020FrP}.}
Determining where turbulent energy is generated, how it is transported, and where it is dissipated is therefore essential for explaining the evolution of the solar wind and the resulting structure of the heliosphere.

The global solar wind is reorganized over the Sun's activity cycle as the distribution of open magnetic flux and solar source regions changes \citep{schwenn2006solar}. 
The \textit{Ulysses} mission revealed a strongly bimodal heliosphere near solar minimum, with fast, relatively steady wind from polar coronal holes (CHs) and slower, variable wind around the equatorial streamer belt; this latitudinal organization gets significantly distorted near solar maximum \citep{mccomas2003grl,ebert2009bulk}. 
At 1 AU, solar-cycle variations have been identified in plasma quantities such as density, magnetic-field strength, and dynamic pressure \citep{crooker1983solar,zhao2018ApJcr2}, while mass and energy fluxes show weaker variability \citep{wang2010relative,le2012solar}. 
Near-Earth turbulence also varies with solar activity, with enhancement of fluctuation amplitudes, Alfv\'enicity, and energy cascade rates at solar maximum \citep{chapman2008solar,manoharan2012three,zhao2018ApJcr2,gautam2024solar}.
These observations show that the solar cycle modulates both the large-scale flow and the character of fluctuations coupled to it.

Launched in 2018, NASA's \textit{Parker Solar Probe} (PSP) mission has accumulated a mature and extensive dataset of in situ observations in the very inner heliosphere \citep{raouafi2023parker}. 
Since PSP has a highly elliptical solar orbit \citep{fox2016SSR}, these data provide an unprecedented opportunity to examine the radial evolution of solar wind in conjunction with solar activity variation. 
PSP's first 25 orbits span the transition from solar minimum (sol-min; \(\sim 2018\)) to maximum (sol-max; \(\sim 2024\); see Fig. \ref{fig:sunspotn}), providing a long-baseline view of the inner heliosphere during markedly different solar and coronal conditions \citep{Usoskin2023LRSP}. 
These observations make it possible to investigate whether solar-cycle trends at 1 AU are already present in the young solar wind, before turbulent fluctuations have fully evolved through expansion, stream interaction, and nonlinear cascade \citep{chen2020ApJS,Chhiber2026ApJ}. 

However, in-situ observations alone do not identify whether the observed variability is produced by local (in situ) evolution or determined by the solar origins of observed wind streams --- a question at the heart of long-standing debates in heliophysics \citep{Viall2020JGR}.
Global numerical models of the corona and solar wind provide the complementary information needed to connect PSP measurements to the Sun \citep{reville2020ApJS,badman2023JGR}.
Models that include Alfv\'en-wave or turbulence-driven acceleration and heating have been developed to describe the coupled evolution of bulk solar wind and its fluctuations \citep{usmanov2000global,van2014alfven,usmanov2018}.
In this study we employ a three-dimensional (3D) global magnetohydrodynamic (MHD) model that includes bulk transport and turbulence modeling in a self-consistent framework \citep{Usmanov2025ApJ}. 
We combine in situ PSP measurements with remote sensing observations of coronal and photospheric structure, employing our global MHD simulations to establish magnetic connectivity between the two regimes, to investigate how solar-cycle--dependent source structure shapes turbulence in the young solar wind. 
We evaluate the mean radial dependence of observed turbulence parameters and investigate departures from these trends, thereby enabling identification of solar wind streams with greater or diminished turbulence energy and Alfv\'enicity relative to the expectation at a given helioradius. 

\added{In the following, Sec. \ref{subsec:solarcycle_var} presents radial distributions of selected turbulence parameters observed by PSP between 0.15 and 0.5 au, highlighting their variation with solar activity and the associated statistical trends. Sec. \ref{subsec:corona_source_region} investigates the relationship between these turbulence properties and their inferred source regions, employing three PSP encounters during low, increasing, and high solar activity as case studies. Sec. \ref{subsec:evolution_turb} examines the modeled evolution of turbulence energy through the corona in three corresponding simulation runs. The results are discussed and summarized in Sec. \ref{sec:discuss}. Several appendices provide additional details on  PSP data analyses, computed turbulence diagnostics, global simulation, and other supplementary results.}
\section{Results}\label{sec:result}
\subsection{Solar-Cycle Variations in the Young Solar Wind Observed by PSP}\label{subsec:solarcycle_var}
\begin{figure*}
    \centering
    \includegraphics[width=0.9\textwidth]{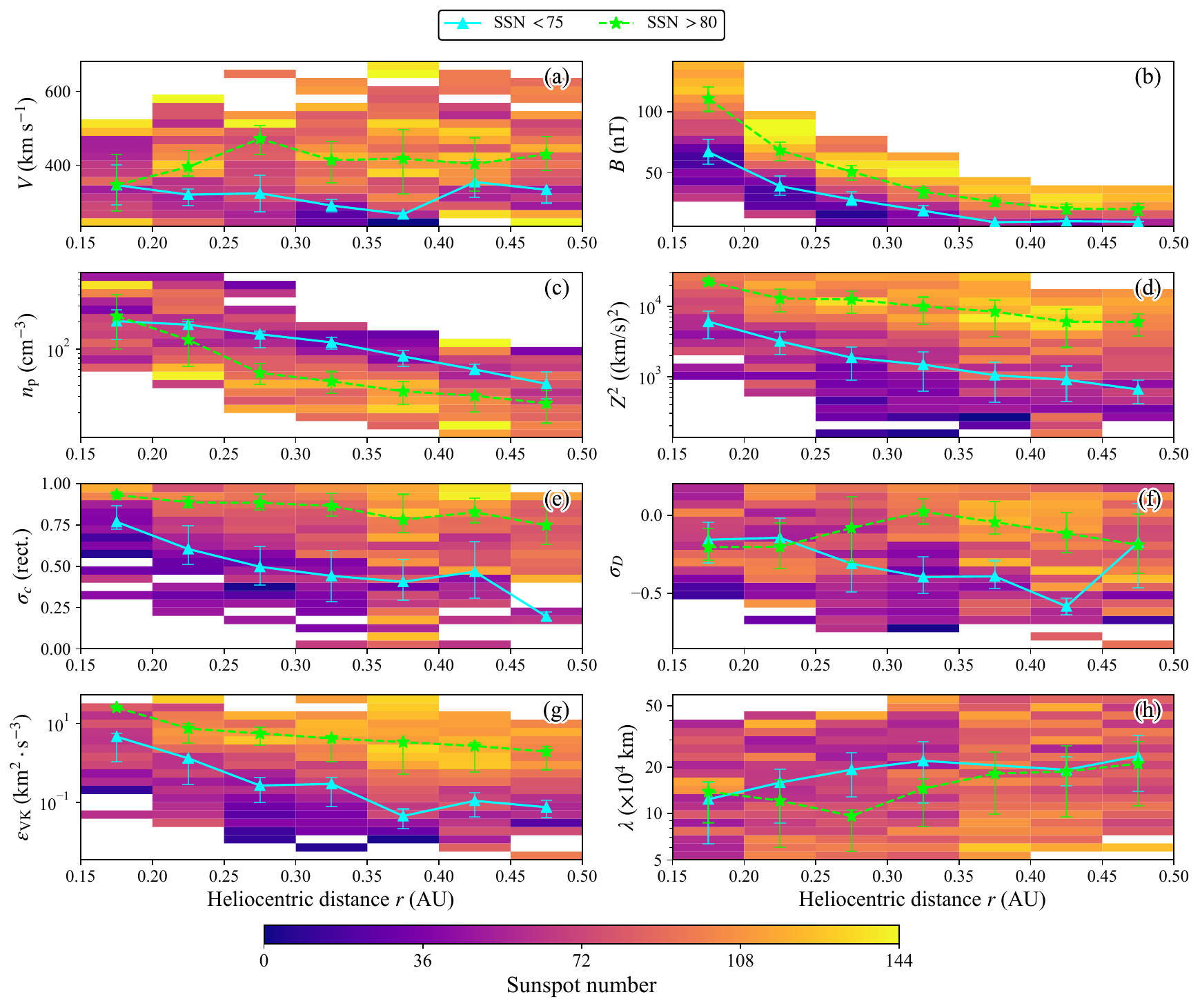}
    \caption{Radial distributions of solar wind properties observed by PSP during orbits 1 to 25, and their variation with SSN.      All quantities represent mean values within non-overlapping 2-hr intervals. 
     The $(r,*)$ plane is divided into bins of width \(0.05\) au, and only bins that contain at least five intervals are shown. The color of each \((r,*)\) bin indicates the mean daily SSN (colorbar) computed across all intervals in the bin. 
     Cyan triangles and green stars indicate conditioned averages in each column (radial bin), computed for SSN \(<75\) and SSN \(>80\), respectively.
    Vertical bars indicate the count-weighted 25-75\% range of the displayed bin means (see App. \ref{app:ssn-binning}).
    Individual panels show: (a) solar wind speed $V$, (b) magnetic field magnitude $B$, (c) proton number density $n_{\mathrm{p}}$, (d) turbulent energy density $Z^2$, (e) sector-rectified cross helicity $\sigma_c$, (f) normalized residual energy $\sigma_D$, (g) von K\'arm\'an turbulent energy-transfer rate $\varepsilon_{\mathrm{VK}} = f(\sigma_c)Z^3/\lambda$, and (h) turbulent correlation length $\lambda$.
         }
    \label{fig:hist2d_var}
\end{figure*}
We employ PSP observations from 2018 October 31 to 2025 November 1, dividing the data into non-overlapping 2-hr intervals, and use the daily sunspot number (SSN) as a reference for solar activity \citep{SILSO_Sunspot_Number}. We restrict this analysis to helioradii (\(r\)) between 0.15--0.5 AU, since PSP data quality and coverage are optimal at smaller \(r\) \citep{SWEAP2020UserGuide}, while below 0.15 AU the validity of Taylor's frozen-in hypothesis \cite[used to convert temporal turbulent correlations to the spatial domain;][]{taylor1938ProcRSL} becomes questionable \citep{chhiber2019psp2}.

For each interval, we compute mean plasma properties and turbulence diagnostics from velocity and magnetic-field fluctuations defined relative to the interval mean. 
We compute (twice) the total incompressible turbulent energy density $Z^2$ as the sum of kinetic and magnetic fluctuation energies; the normalized, sector-rectified cross helicity $\sigma_c$, which measures the correlation of velocity and magnetic fluctuation vectors; and the normalized residual energy $\sigma_D$,  a measure of the difference between kinetic and magnetic fluctuation energies \citep{bruno2013LRSP}. Note that \(\sigma_c\) and \(\sigma_D\) are metrics of the Alfv\'en-wave-like character of the fluctuations, with \(\sigma_c=\pm1\) and \(\sigma_D=0\) for a linear Alfv\'en wave \citep{parashar2020ApJS}.\footnote{While these properties are known to have a scale-dependent character, our choice of averaging interval (2 hr) implies that the computed values represent the energy-containing range of turbulence scales \citep{pope2000book,chen2020ApJS}.}

We also estimate the correlation length of magnetic fluctuations ($\lambda$), which is used both as an observed energy-containing scale \citep{pope2000book} and as the outer-scale length to compute the von K\'arm\'an energy decay rate (denoted hereafter as energy transfer rate), modified for MHD: $\varepsilon_{\mathrm{VK}}= f(\sigma_c)Z^3/\lambda$ \citep{karman1938prsl,hossain1995PhFl,matthaeus2004grl}. 
\(\varepsilon_\text{VK}\) estimates the rate of energy transfer from the large energy-containing scales of turbulence to the inertial range, and can provide a good approximation to the inertial-range cascade rate \citep{bandyopadhyay2018epsilon,bandyopadhyay2020ApJS_cascade,Wu2022ApJ} and the heating rate of the plasma \citep{cranmer2009ApJ,Usmanov2025ApJ}.
The factor \(f(\sigma_c)\) accounts for the modulation of the turbulent cascade by finite \(\sigma_c\).
Full definitions and data processing details are given in App. \ref{app:psp-data} and \ref{app:turbulence-diagnostics}. 

Figure~\ref{fig:hist2d_var} shows radial distributions of these quantities, with the colormap indicating SSN. Average radial trends conditioned on the SSN are also displayed, revealing the distinct absolute values and radial evolution during solar min and max. Higher-SSN intervals generally have faster speeds, stronger magnetic fields, and greater \(Z^2\), \(\sigma_c\), and energy-transfer rates.
As $f(\sigma_c)$ tends to vanish  when $|\sigma_c|\rightarrow1$, the larger cross helicity at high SSN would, by itself, weaken nonlinear interactions \citep{matthaeus2004grl,Chhiber2026ApJ}; the simultaneous enhancement of $\varepsilon_{\rm VK}$ therefore indicates that the increase in fluctuation amplitude
 outweighs this Alfv\'enic suppression. 
The comparatively weak solar-activity dependence of $\lambda$ provides little compensation for this increase.
The residual energy $\sigma_D$ also shifts, for the most part, toward less negative values during higher-activity periods, which, together with the increase in \(\sigma_c\), indicates a shift towards more Alfv\'enic turbulence. 

The faster speeds and lower proton densities are consistent with the turbulence trends \citep{bruno2013LRSP,Chhiber2026SpaceWeather}, giving an overall impression that fast, Alfv\'enic, coronal-hole wind is being sampled with greater frequency in the ecliptic region during periods of high solar activity.
The radial trends in cross-helicity are strikingly different during low and high activity: sustained highly-imbalanced \(\sigma_c\) values at \(r\sim\) 0.15 -- 0.5 au during sol-max, in contrast to rapid reduction of Alfv\'enicity with \(r\) during sol-min.

These trends indicate that the solar cycle modulates both the large-scale plasma environment and the turbulent state of the very inner heliosphere. The qualitative trends with SSN are consistent with those observed at 1 AU \citep{zhao2018ApJcr2,gautam2024solar}, indicating that the patterns of variability observed in near-Earth turbulence are already established near the Sun, \added{even as radial evolution appears to accentuate the differences between sol-min and sol-max (Fig. \ref{fig:hist2d_var})}. Moreover, our results show that the magnitude of variability produced by solar activity is of a level comparable to the variation with \(r\).

To further elucidate these trends, the next section presents an analysis of the solar connectivity of observations from \added{three PSP orbits that sample periods of varying solar-activity levels.}

\subsection{\added{Solar Source-Region} Influence on Young Solar-Wind Turbulence}\label{subsec:corona_source_region}

\added{The three PSP encounters examined here are E6, E14, and E22, corresponding to Carrington Rotation (CR) numbers 2235, 2265, and 2293. We employ three corresponding simulations runs of the \cite{Usmanov2025ApJ} global model to establish magnetic connectivity between the PSP trajectory and the coronal base (see App. \ref{app:mhd-connectivity}), thereby inferring the source regions of observed streams. The three encounters and solar configurations are samples from low, increasing, and high solar-activity epochs (see Fig. \ref{fig:sunspotn}), and serve as preliminary case studies rather than a comprehensive representation of the three activity levels.} The three simulations use the same governing equations, turbulence closures, and coronal-base turbulence parameters; their difference lies in the observed map of radial magnetic field imposed at the coronal boundary. See App. \ref{app:mhd-connectivity} for details. 

For each valid 2-hr interval between 0.15--0.5 au, PSP's position is computed at the interval midpoint, and the corresponding magnetic field-line is traced through the model to the coronal-base boundary at $1\ R_{\odot}$. \added{Below, we overlay traced footpoints on extreme-ultraviolet (EUV) images of the solar corona from the Solar Dynamics Observatory's Advanced Imaging Assembly \citep[SDO/AIA;][]{Lemen2012SoPh}, which contain information about large-scale coronal structures associated with the magnetic field, such as active regions (ARs) and coronal holes \citep{DelZanna2018LRSP_Solar_UV_Xray}. 
Source regions are evaluated with respect to \(Z^2\) and \(\sigma_c\) levels observed at PSP; since both quantities systematically decrease between 0.15 and 0.5 au (Fig. \ref{fig:hist2d_var}), PSP intervals are assigned a normalized (or detrended) measure based on empirical fits to the observed radial trends. These empirical fits are computed separately for low, increasing, and high activity epochs. The detrended measures are the normalized turbulence energy $R_{Z^2}=Z^2/[Z^2_{\rm fit}(r)]$, and the excess/deficit cross helicity \(\Delta\sigma_c=\sigma_c -\sigma_{c,\text{fit}}(r)\); they measure the degree of departure from the expected value at a given \(r\). See App. \ref{app:relative-turbulence-selection} for details of the fitting procedure and the three activity epochs used.} 
\subsubsection{PSP E6; CR 2235; Low Solar Activity}
\begin{figure}
    \centering
    \includegraphics[width=0.5\textwidth]{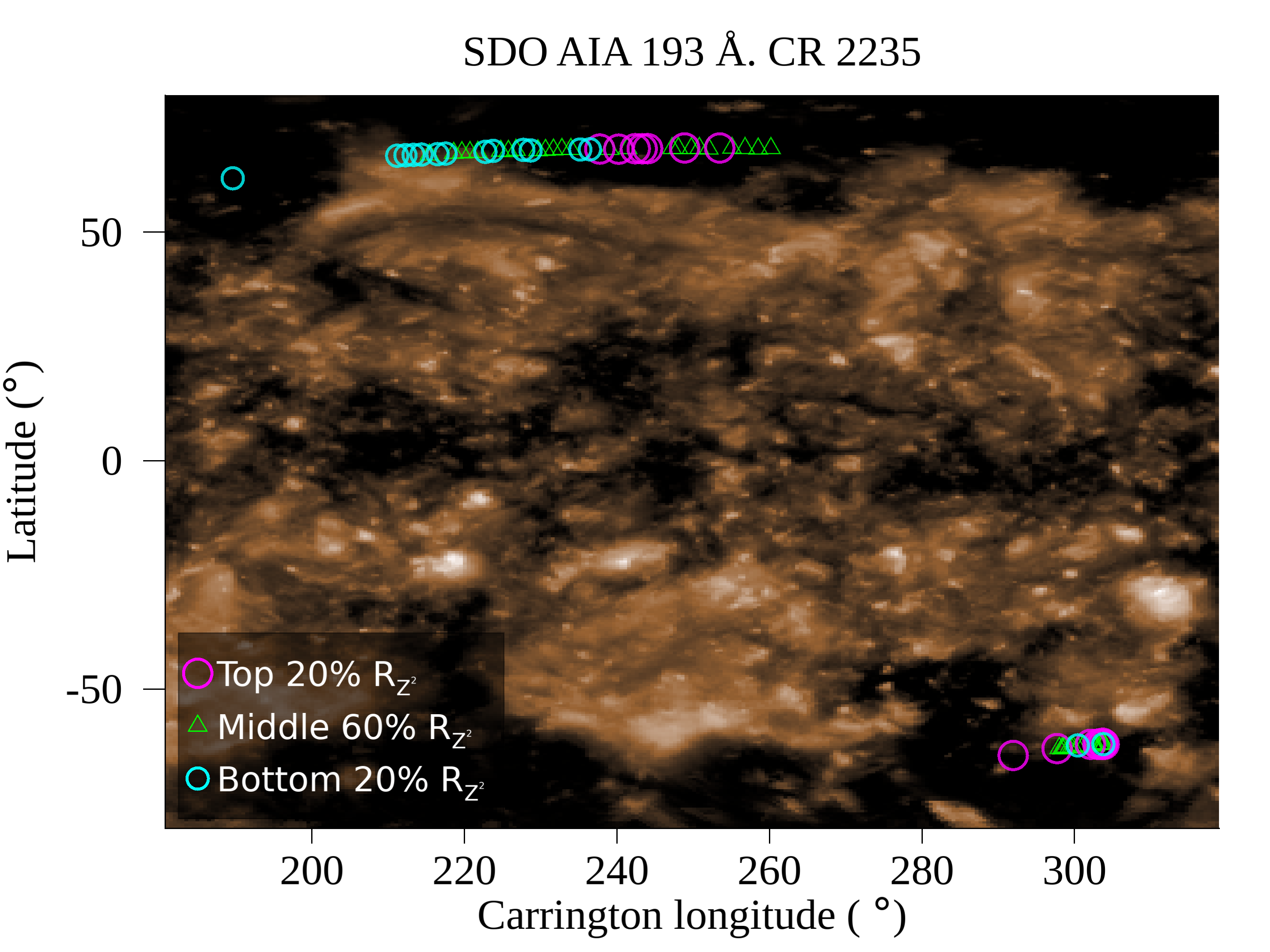}
    \includegraphics[width=0.5\textwidth]{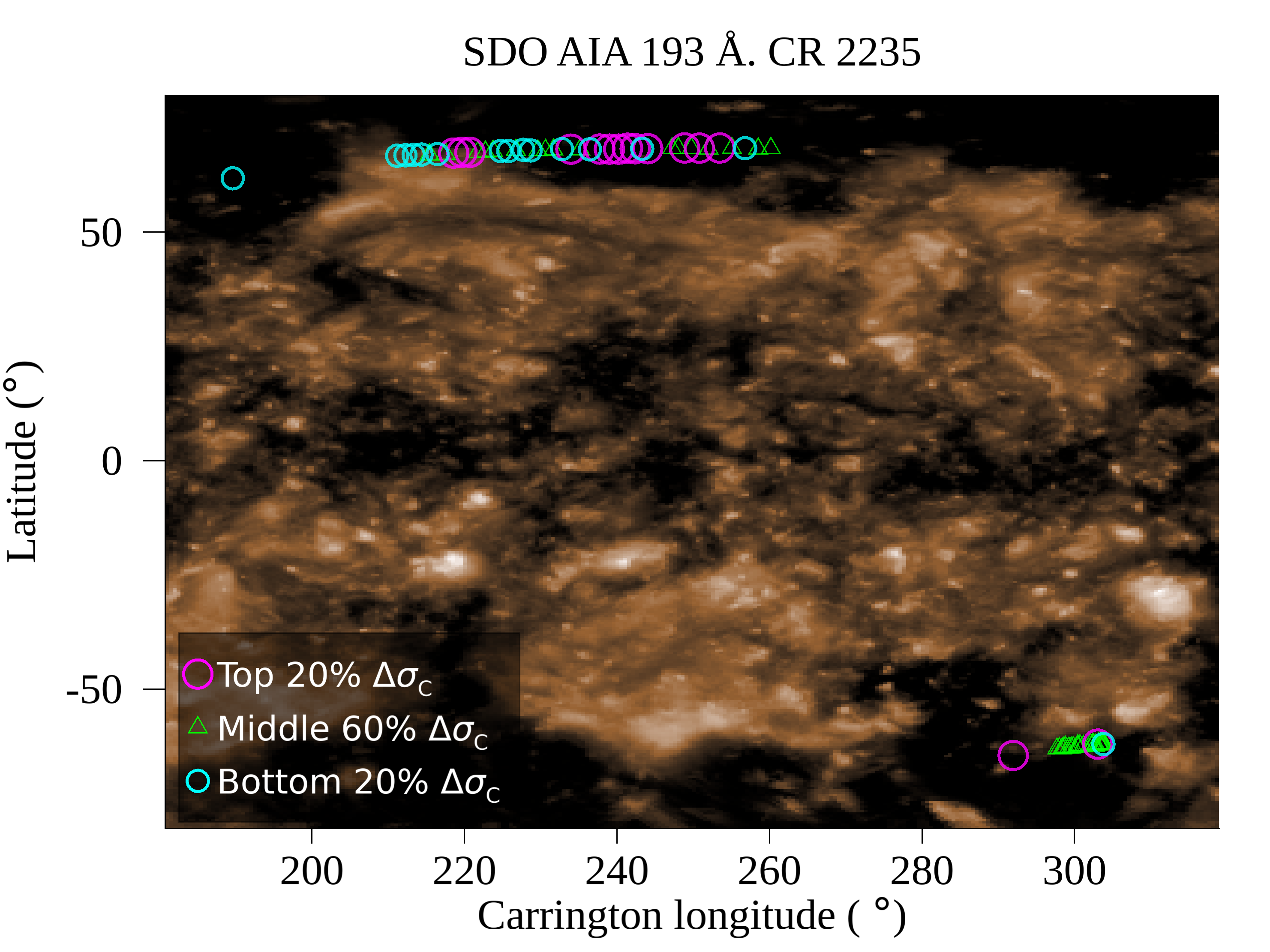}
    \caption{
    Inferred low-coronal sources of PSP's E6 observations of the solar wind with varying levels of (top) turbulent energy and (bottom) cross helicity. Circles mark footpoints of SW streams, overlaid on a background cropped from an SDO/AIA 193 \AA~EUV synoptic map for CR 2235. Axes indicate heliographic latitude and Carrington longitude. SW intervals are classified by their detrended turbulence-energy density $R_{Z^2}$ and cross helicity \(\Delta\sigma_c\).
    Large (magenta) and small (cyan) circles mark sources of the upper and lower 20\% of $R_{Z^2}$, respectively, while green triangles mark the remaining 60\%. See also Fig \ref{fig:trace_Br}(a).
    }
    \label{fig:2235_aia_ftpt}
\end{figure}
Footpoints successfully traced from PSP's trajectory in E6 (Sep--Oct 2020) to the coronal base are overlaid on 193 \AA~EUV synoptic maps in Fig.~\ref{fig:2235_aia_ftpt}. E6 intervals with the upper and lower 20\% of $R_{Z^2}$ define the high and low turbulence-amplitude subsets, respectively; remaining intervals form the middle 60\%. An analogous distribution is employed for \(\Delta\sigma_c\). Distinct symbols that mark the footpoints in Fig. \ref{fig:2235_aia_ftpt} correspond to these three subsets, discussed further below.

For E6, the footpoints are confined predominantly to two high-latitude bands near the solar poles (Fig.~\ref{fig:2235_aia_ftpt}). In the top panel, upper \(R_{Z^2}\) subsets (large magenta circles) tend to come from dark regions in the interior of these polar CHs, while the lower \(R_{Z^2}\) intervals (small cyan circles) tend to originate near the boundary between dark and bright regions in the EUV map, corresponding to a transition from unipolar CHs to regions with slightly weaker magnetic field, bordering mixed-polarity structures [see Figs. \ref{fig:AIA_intensity}(a) and \ref{fig:trace_Br}(a)]. 

The source-region organization of low/high Alfv\'enicity intervals (Fig. \ref{fig:2235_aia_ftpt}, bottom) is similar to that of \(R_{Z^2}\), suggesting that solar sources exert influence on both the amplitude and the Alfv\'enic imbalance of solar-wind turbulence \cite[see also][]{Borovsky2019JGR,DAmicis2025AA,Alterman2025ApJ}. This similarity is also indicative of a positive correlation between fluctuation amplitude and Alfv\'enicity \citep[see also][]{Chhiber2026SpaceWeather}. 

\added{In App. \ref{app:mhd-connectivity}, panels (a) and (d) of Fig. \ref{fig:AIA_intensity} show scatterplots of AIA 193-\AA\ intensity vs \(R_{Z^2}\) and \(\Delta \sigma_c\), respectively, for E6. This quantitative assessment indicates a weak-to-moderate anticorrelation between brightness and the turbulence metrics; the correlation with magnetic field magnitude \(|B_r|\) at the footpoints is weaker still.} 

\begin{figure}
    \centering
    \includegraphics[width=0.5\textwidth]{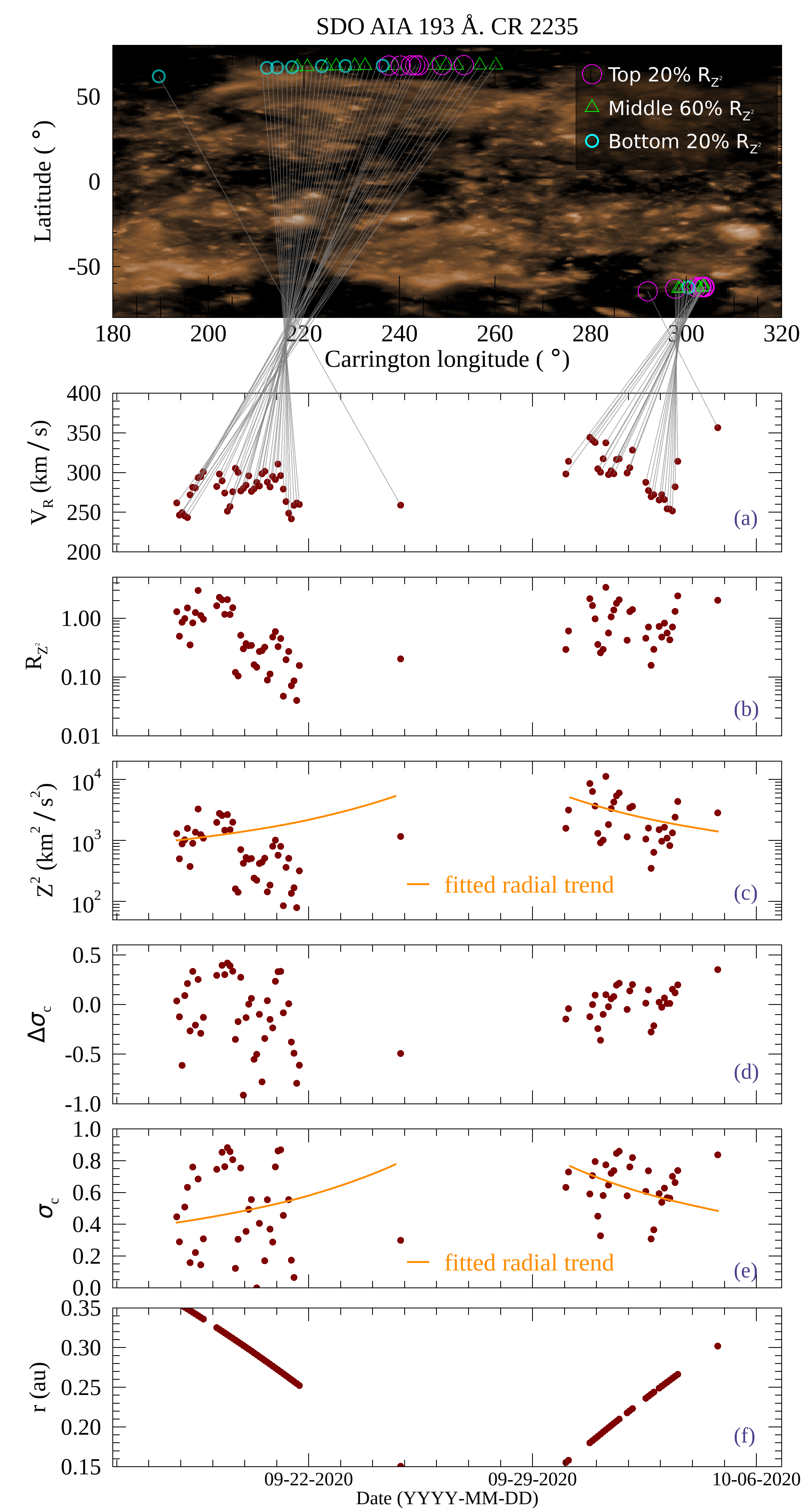}
    \caption{
    Top panel shows EUV map marking inferred sources of solar wind observed by PSP during E6, following Fig. \ref{fig:2235_aia_ftpt}. Panels (a)--(f) show timeseries of radial wind speed (\(V_R\)), \(R_{Z^2}\), \(Z^2\), \(\Delta\sigma_c\), \(\sigma_c\), and helioradis of PSP during E6, from top to bottom. Minor tick-marks on date axes are 1 day apart. Each brown dot represents an average over a 2-hr interval. Orange curves in panels (c) and (e) show fitted radial trends for \(Z^2\) and \(\sigma_c\), respectively, for the sol-min epoch. \(R_{Z^2}>1\) and \(\Delta\sigma_c>0\) indicate turbulence amplitude and \(\sigma_c\) values that are greater than the respective radial trends, while \(R_{Z^2}<1\) and \(\Delta\sigma_c<0\) indicate the opposite case. Grey lines that span panel (a) and the EUV map connect PSP intervals to their respective footpoints. 
    }
    \label{fig:timeser1}
\end{figure}
\added{To further elucidate the source connectivity, Fig. \ref{fig:timeser1} shows timeseries of selected quantities observed by PSP in E6, and explicitly represents the connection between shown time intervals and their respective footpoints on the EUV map. The fitted radial trends of \(Z^2\) and \(\sigma_c\) for sol-min are also shown, demonstrating the impact of the detrending step and its crucial role in identifying variability in turbulence parameters beyond what is expected from radial evolution.\footnote{{See also Fig. \ref{fig:Z2_radial_dependence}, which shows the radial fit explicitly as a function of \(r\).}} The general transition from large/medium \(\RZ\) to smaller values in the time period of \(\sim\) 17 -- 22 Sep is apparent as PSP's source connection shifts from darker to brighter source regions on the EUV map.}
\subsubsection{PSP E14; CR 2265; Increasing Solar Activity}
\begin{figure}
    \centering
    \includegraphics[width=0.5\textwidth]{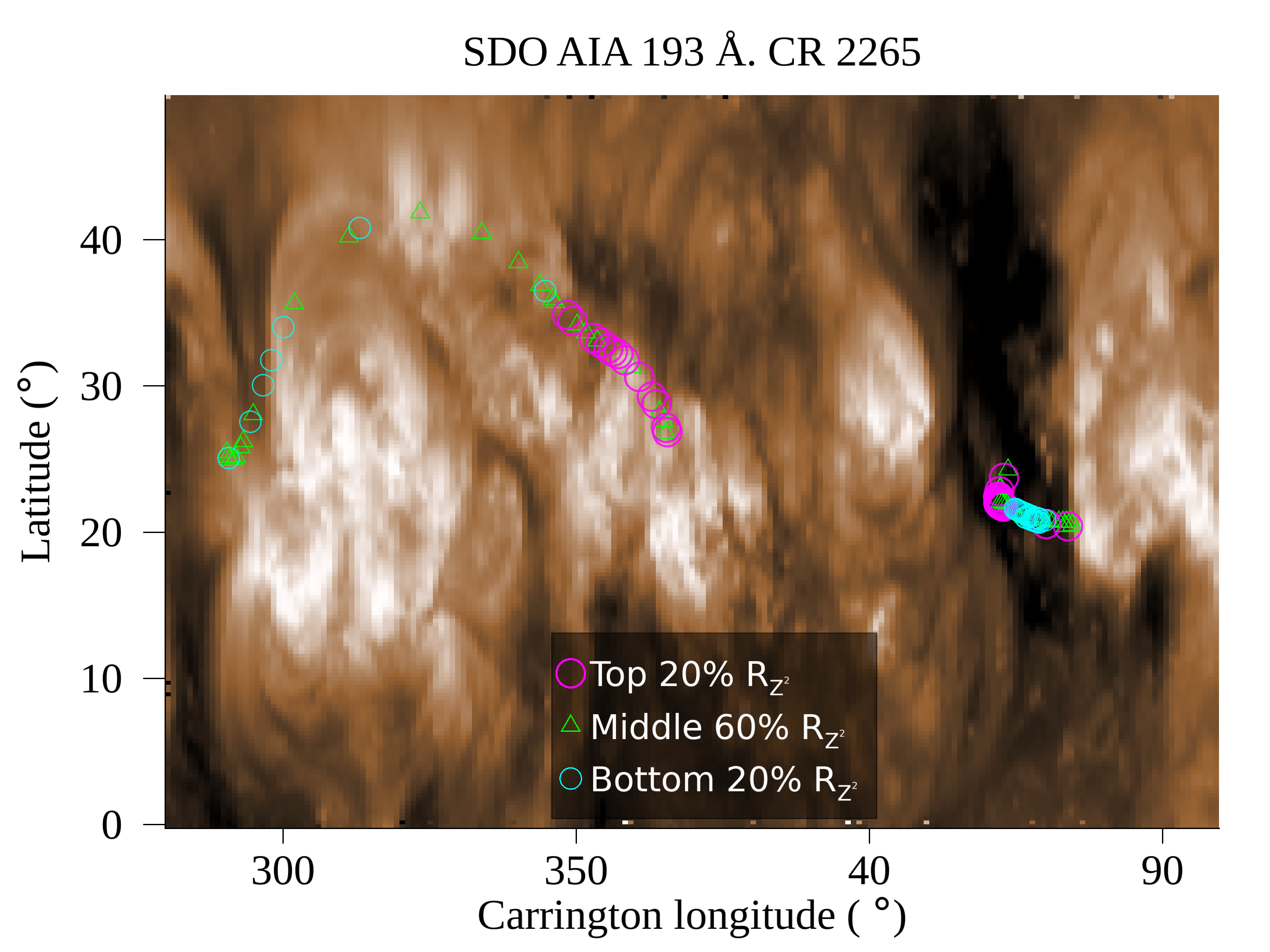}
    \includegraphics[width=0.5\textwidth]{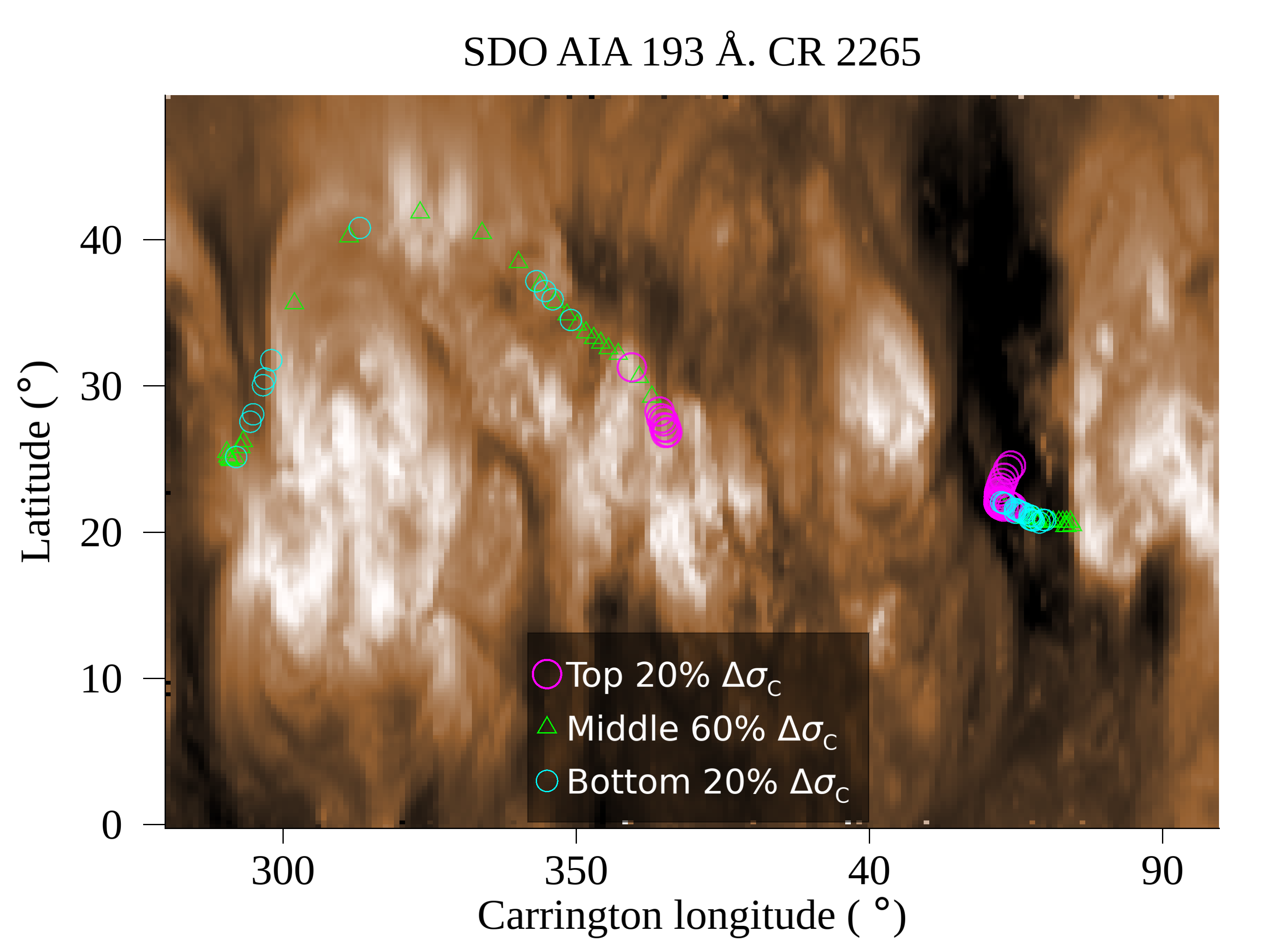}
    \caption{
    Inferred low-coronal sources during CR 2265 for PSP E14, following a format analogous to that of Fig. \ref{fig:2235_aia_ftpt}. See also Fig. \ref{fig:trace_Br}(b). Note that the periodicity of Carrington longitude has been employed in order to shift values between 0\degree\ and \(\sim100\)\degree\ to the right side of the image.
    }
    \label{fig:2265_aia_ftpt}
\end{figure}

\added{Fig. \ref{fig:2265_aia_ftpt} shows the footpoint distribution for PSP's E14 (Nov--Dec 2022), corresponding to CR 2265, a period of increasing solar activity (see Fig. \ref{fig:sunspotn}). In this case, it is difficult to discern a pattern relating \(\RZ\) and \(\Delsig\) levels to source structure: the large CH on the right side of the EUV map is the inferred source of both magenta and cyan intervals, while the brighter regions on the left side are also somewhat mixed. The scatterplot of turbulence parameters against AIA intensity (Fig. \ref{fig:AIA_intensity} in App. \ref{app:mhd-connectivity}) also indicates a negligible or weak correlation. However, the footpoints do show clear sequential patterns that suggest a somewhat ordered transition from magenta-to-green-to-cyan.}\footnote{\added{We note a moderately dark region in Fig. \ref{fig:2265_aia_ftpt} just above the magenta footpoints between \(\sim340\)\degree\ and 360\degree\ longtitude, which is possibly obscured by brighter AR loops due to line-of-sight projection effects \citep{,Kirk2009SoPh,ODwyer2010AA}. See also Fig. \ref{fig:trace_Br}(b), which indicates a unipolar magnetic corridor associated with the sequence of magenta-to-green footpoints moving from \(\sim 360\)\degree\ to 320\degree\ longitude.}}

\begin{figure}
    \centering
    \includegraphics[width=0.5\textwidth]{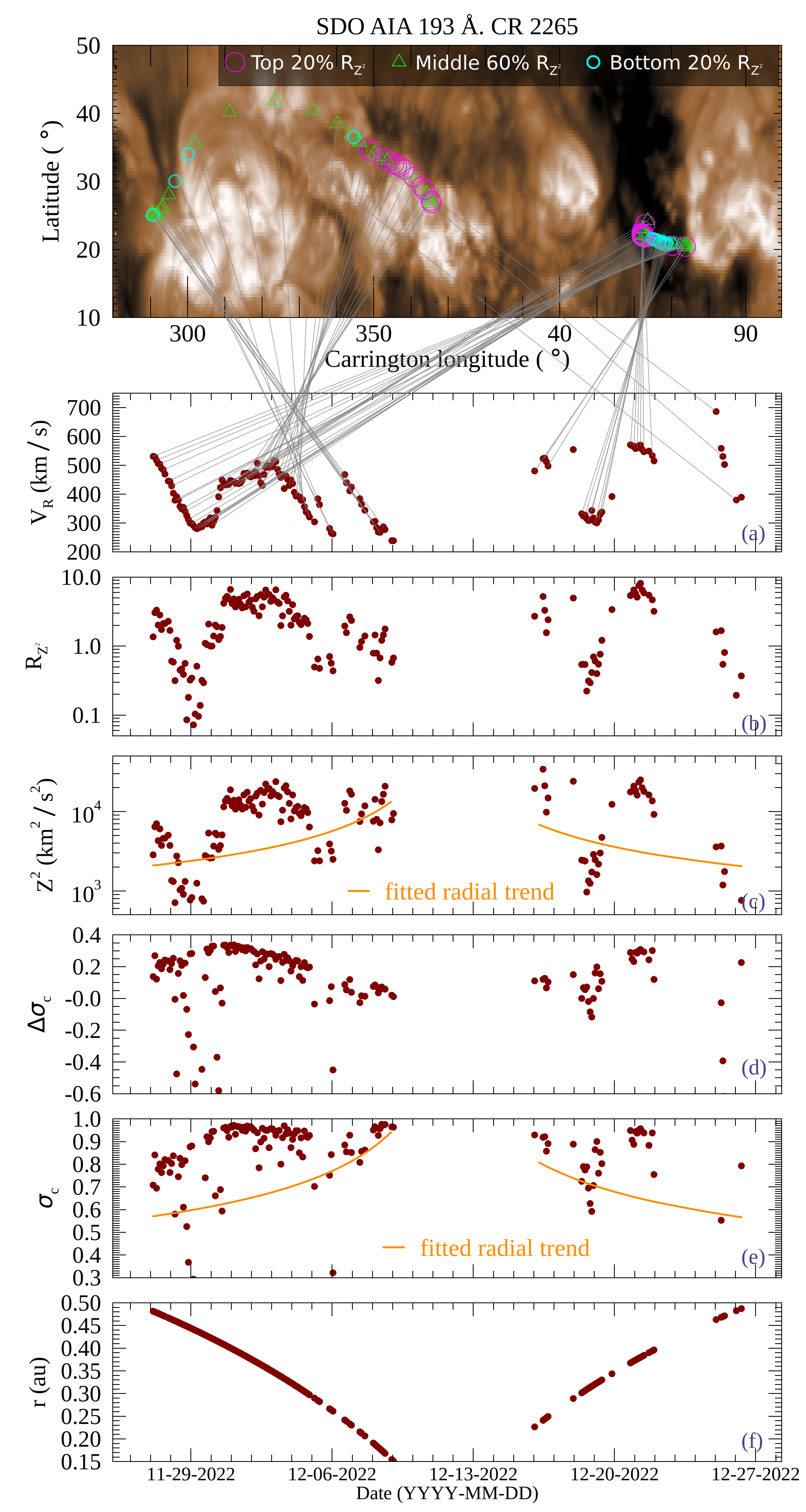}
    \caption{
     Timeseries of selected quantities observed by PSP during E14, with a representation of the connection between PSP time-intervals and their inferred source-region footpoints on the CR 2265 EUV map. The format is analogous to that of Fig. \ref{fig:timeser1}. Radial trends shown in panels (c) and (e) are obtained from empirical fits to observations during the increasing solar-activity epoch.
    }
    \label{fig:timeser2}
\end{figure}

\added{These sequential patterns are clearer in the timeseries (Fig. \ref{fig:timeser2}), where a few instances of relatively smooth transition from medium/fast to slow wind are evident.}\footnote{\added{At these relatively small heliodistances, fast wind streams may not have reached their asymptotic, fully-accelerated speeds \citep{Rivera2024Sci,Usmanov2025ApJ}.}} \added{This case also demonstrates, more strikingly than Fig. \ref{fig:timeser1}, the variability of turbulence properties between streams. The contrasting trends in \(\sigma_c\) and \(\Delsig\) between 5 to 9 Dec are especially interesting, and demonstrate clearly the impact and utility of the detrending step.}
\subsubsection{PSP E22; CR 2293; High Solar Activity}
\begin{figure}
    \centering
    \includegraphics[width=0.5\textwidth]{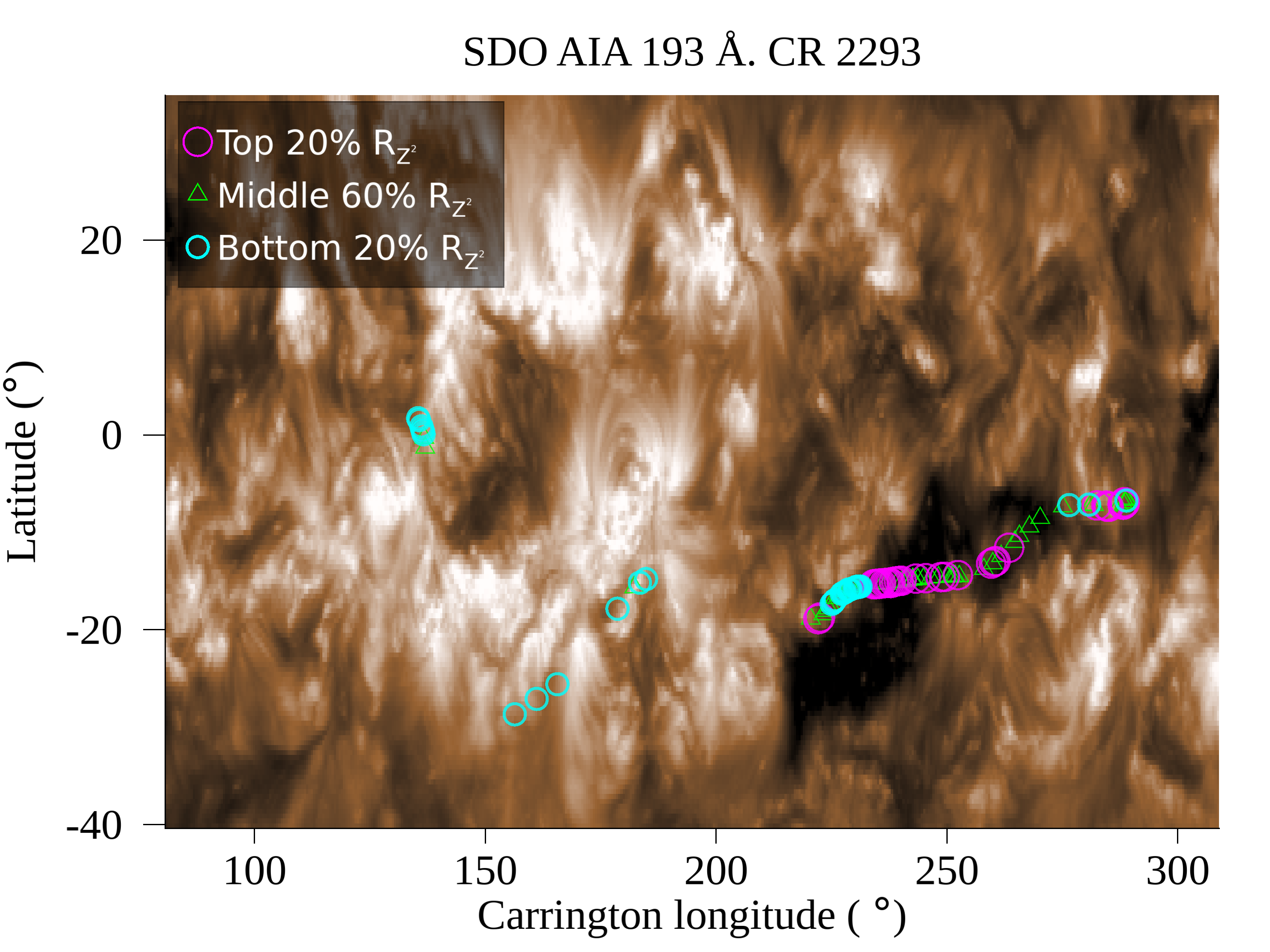}
    \includegraphics[width=0.5\textwidth]{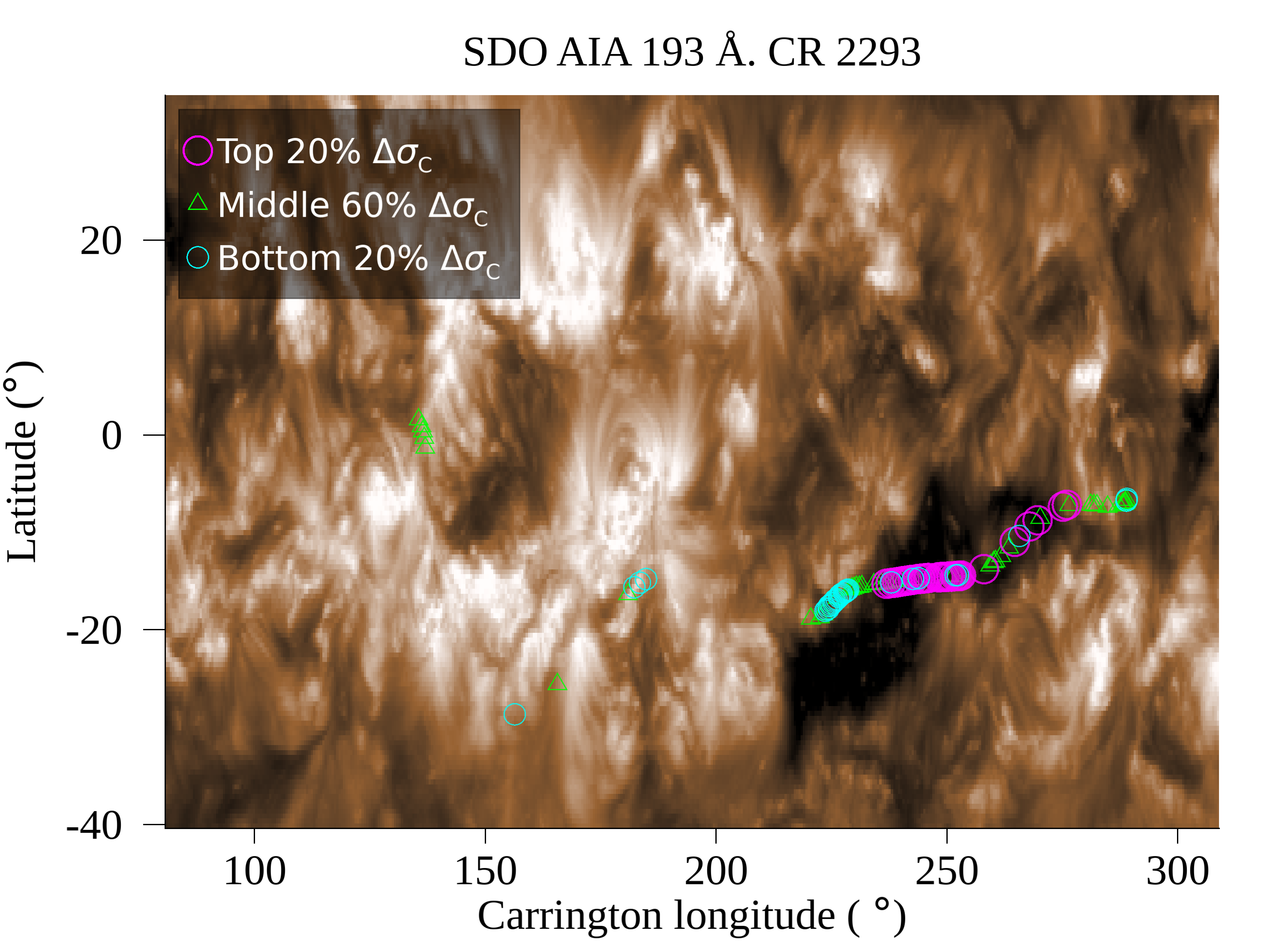}
    \caption{Inferred low-coronal sources during CR 2293 for PSP E22, following a format analogous to that of Fig. \ref{fig:2235_aia_ftpt}. See also Fig \ref{fig:trace_Br}(c).
    }
    \label{fig:2293_aia_ftpt}
\end{figure}
Our final case is PSP's E22, a period of high solar activity (see Fig. \ref{fig:sunspotn}). Footpoints are depicted in Fig. \ref{fig:2293_aia_ftpt} in the established format. This case presents a striking differentiation between sources of low and high \(R_{Z^2}\) (and \(\Delsig\)): the upper-$R_{Z^2}$ subset is clustered within a dark EUV corridor consistent with a CH. In contrast, the lower $R_{Z^2}$ footpoints extend into the surrounding bright and highly-structured EUV environment, and notably within the AR complex between \(\sim 120\degree\)~and \(200\degree\)~longitude. 
These low-$R_{Z^2}$ footpoints generally do not coincide with the brightest EUV cores; instead, they lie preferentially along diffuse AR peripheries and in interstitial lanes between adjacent bright structures. The multipolar topology of these regions is evident in the photospheric magnetogram shown in Fig. \ref{fig:trace_Br}(c). The source region organization of low/high Alfv\'enicity intervals is rather similar to that of \(R_{Z^2}\).

\added{The ordering of \(R_{Z^2}\) with footpoint EUV intensity is made quantitatively explicit in Figs. \ref{fig:AIA_intensity}(c) and (f) of App. \ref{app:mhd-connectivity}, where we report moderate anti-correlation values. In contrast, the correlation with source \(|B_r|\) is again negligible or weak.}

\begin{figure}
    \centering
    \includegraphics[width=0.5\textwidth]{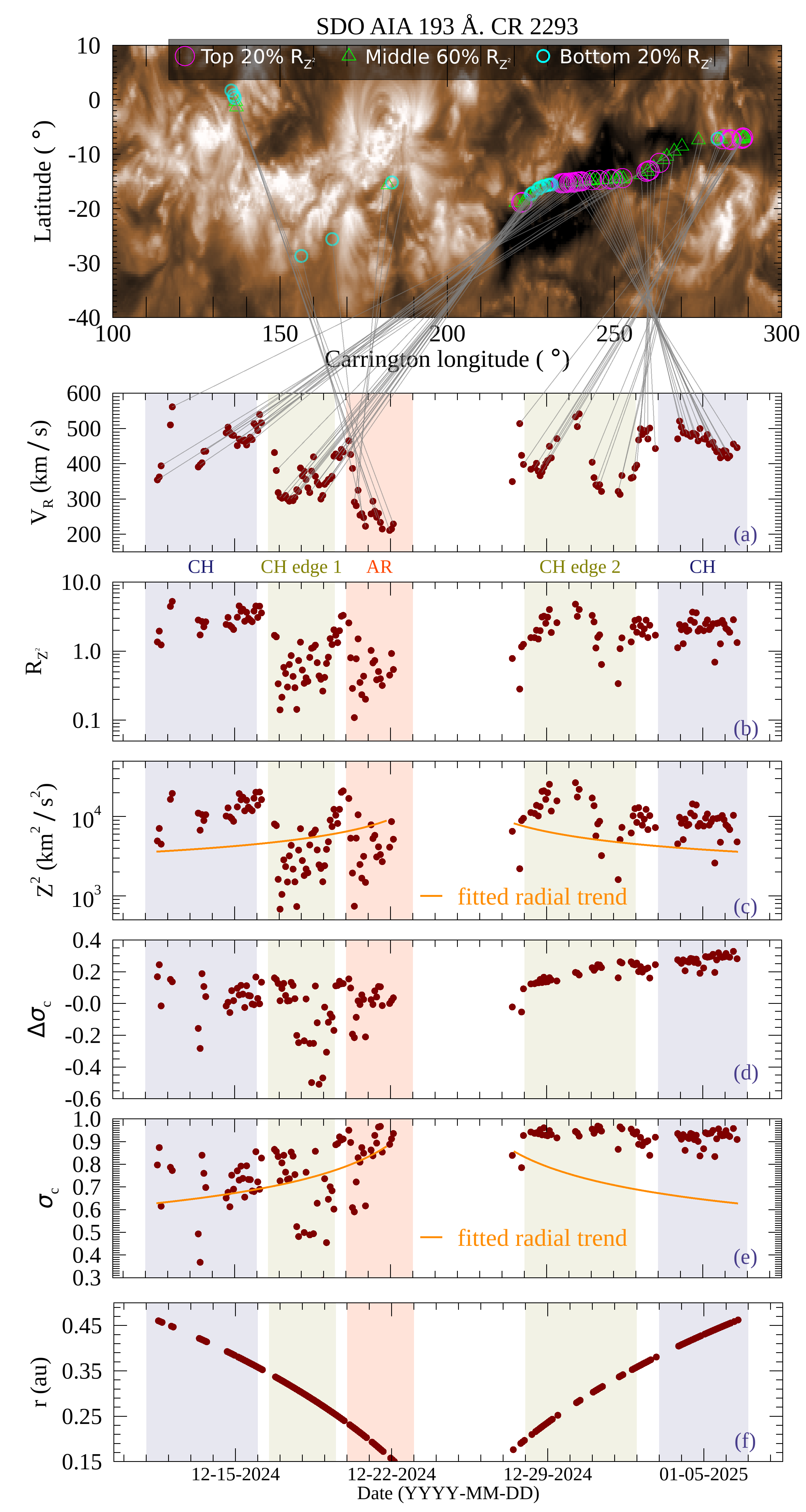}
    \caption{
    Top: Timeseries of selected quantities observed by PSP during E22, with a representation of the connection between PSP time-intervals and their source-region footpoints on the CR 2293 EUV map. The format is analogous to that of Fig. \ref{fig:timeser1}. Radial trends shown here are obtained from empirical fits to observations from the high solar-activity epoch. Source regions of different segments of the timeseries (indicated via varying background shade) are stated in the text between panels (a) and (b): central region of coronal hole (CH; lavender shade), AR complex (peach shade), and CH edges (olive green shade).
    }
    \label{fig:timeser3}
\end{figure}
\begin{figure*}
    \centering
    \includegraphics[width=0.8\linewidth]
    {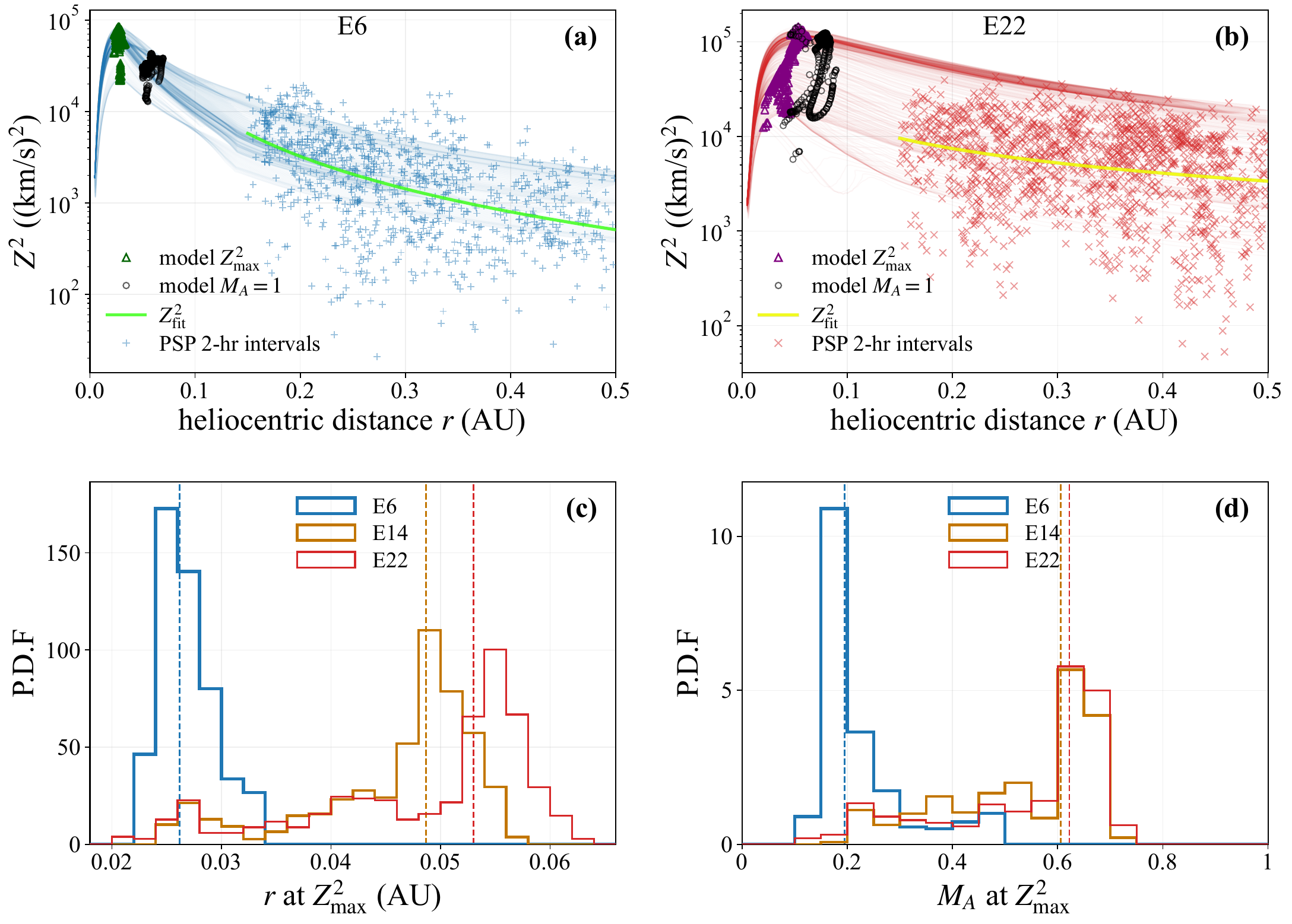}
    \caption{
    \textit{Top}: Radial evolution of modeled turbulent energy along magnetic field lines connected to PSP during (a) low and
    (b) high solar activity, corresponding to E6 and E22, respectively.
    The faint curves show $Z^2$ sampled along individual field lines traced from the PSP trajectory to the model's inner boundary.
    Triangles mark the maximum modeled turbulent-energy density, $Z^2_{\max}$, along each field line, while black circles mark the $M_A=1$ location along the same field line.
    Blue `+' and red `\(\times\)' symbols show PSP observations from the low and high activity intervals between 0.15 AU and 0.50 au. 
    Bright green and yellow curves show radial trends obtained from empirical fits to $Z^2$, for low and high activity PSP datasets, respectively (see App. \ref{app:relative-turbulence-selection}).
    \newline
    \textit{Bottom}: (c): PDFs of the heliocentric distance at which $Z^2$ reaches its maximum along each field line (bin size = 0.002 au).
    (d): PDFs of the modeled Alfv\'en Mach number $M_A$ evaluated at the same $Z^2_{\max}$ locations (bin size = 0.05).
    Blue, golden, and red lines represent E6, E14, and E22.
    Dashed vertical lines mark the median of each distribution.
    }
    \label{fig:Z2_radial_dependence}
\end{figure*}

Further insight can be gained by examining the connectivity via the timeseries (Fig. \ref{fig:timeser3}). To highlight the separation of inferred source regions, different segments of the timeseries are indicated via changing color shading of the background. Initially, PSP is connected to the interior of the low-latitude CH, where it measures relatively fast wind, high \(R_{Z^2}\), and excess \(\sigma_c\). This period corresponds to the cluster of nearly-continuous magenta footpoints. The magnetic connection moves toward smaller longitudes, shifting from the CH interior towards the AR environment from 16 to 20 Dec, while passing through the left edge of the CH (edge 1). The peach-shaded segment of the timeseries is connected to the AR complex, with low speeds and \(R_{Z^2}\) levels, and small or negative \(\Delta\sigma_c\). On PSP's return from perihelion (\(\sim 0.1\) AU), it is connected to the region on the right of the CH (edge 2), where the wind properties progress towards faster speeds and greater \(R_{Z^2}\) and \(\sigma_c\) as the footpoints move ``left''-ward and once again enter the CH. 

It is noteworthy that the low and high fluctuation-amplitude intervals indicate \(Z^2\) values an order-of-magnitude apart (\(R_{Z^2}\) of \(\sim0.3\) vs \(\sim3\)). We also observe intriguing enhancements in \(V_R\) and \(R_{Z^2}\) at the interfaces between timeseries segments with different source-region connectivity; these hint at local acceleration at the boundaries between wind streams originating in different sources \citep{Richardson2018LRSP}, possibly driven by shear-driven turbulence generation at such interfaces \citep{roberts1992jgr,deforest2016ApJ828,ruffolo2020ApJ,Usmanov2025ApJ}.

In the next section we examine the \textit{modeled} $Z^2$ evolution along the same traced flux tubes, in an attempt to ``fill the gap'' between the in-situ measurements above 0.15 AU and the remote solar observations.
\subsection{Modeled Evolution of Turbulent Energy through the Corona}\label{subsec:evolution_turb}

We sample the modeled \(Z^2\) along every successfully-traced magnetic field-line in the simulations for low, increasing, and high solar-activity cases. Each profile extends from the model inner boundary at $r=R_{\odot}$ to the corresponding point along the PSP trajectory. \added{We only show the low and high activity cases in Figs. \ref{fig:Z2_radial_dependence} (a) and (b), since the radial evolution of the latter is very similar in the increasing-activity case.} In each case, $Z^2$ evolves non-monotonically along the traced flux tubes, following a pattern consistent with several previous modeling studies \citep[e.g.,][]{verdini2009ApJ,chhiber2019psp1,Usmanov2025ApJ}: it increases rapidly above the solar surface, reaches a maximum in the middle--outer corona \citep{Seaton2023SoPh}, and subsequently decreases with \(r\).  
The turbulence measured by PSP above 0.15 au therefore samples the outward-decreasing portion of the modeled profiles.

The PSP measurements are included as an observational comparison over $0.15\leq r < 0.50$ AU in Fig.~\ref{fig:Z2_radial_dependence} (a) and (b). As stated elsewhere, we only analyzed PSP data above 0.15 AU in view of complications relating to Taylor's hypothesis and SWEAP data.
At overlapping heliocentric distances, however, the observed values occupy the same broad range as the modeled profiles and reproduce their overall outward decrease. 
\added{The empirically-fitted trends in $Z^2$ lie approximately within the center of the scattered observational data points, and closely follow the radial trends of the modeled magnetic field lines.}
Such agreement in turbulent energy between our global MHD model and PSP observations has been observed in previous work that examined other time periods \citep{bandyopadhyay2020ApJS_cascade,chhiber2021ApJ_psp,Usmanov2025ApJ}.

The location of the modeled maximum of \(Z^2\) changes substantially with solar activity, as indicated in the probability density functions (PDFs) shown in Fig. \ref{fig:Z2_radial_dependence}(c). 
In the low activity case, $Z^2_{\max}$ is concentrated within a relatively narrow radial range near 0.025-0.03 AU. 
In the high-activity case, the distribution is considerably broader and displaced outward to approximately 0.05-0.06 AU [Fig.~\ref{fig:Z2_radial_dependence}(c)]. The Alfv\'en Mach number evaluated at $Z^2_{\max}$ provides additional context for this shift; $M_A = V/V_A$, where \(V_A=B/\sqrt{\mu_0\rho}\).
The maximum of \(Z^2\) occurs in sub-Alfv\'enic flow in both cases, but the high-activity distribution extends to substantially larger $M_A$ and lies closer to the modeled Alfv\'en surface [Fig.~\ref{fig:Z2_radial_dependence} (c) and (d)]. In contrast, in the sol-min case the maxima of $Z^2$ occur deeper in the sub-Alfv\'enic corona and occupy a more compact range of both radius and $M_A$.
\added{The intermediate case (corresponding to E14) is also shown in panels (c) and (d). The SSN during E14 is substantially higher than in E6, but still lower than during E22, as shown in Fig.~\ref{fig:sunspotn}.
As expected, the distributions of helioradius $r$ and Alfv\'en Mach number $M_A$ associated with the maximum of $Z^2$ for E14 lie between those for E6 and E22, but are quite close to the E22 case.
The E14 results therefore provide additional evidence for the significant influence exerted by solar activity on the spatial distribution and radial evolution of inner-heliospheric turbulence.
}

\section{Discussion \& Conclusions}\label{sec:discuss}
\added{We have investigated two questions in this study -- does the solar-activity cycle modulate the turbulent properties of the young solar wind? And does variation in solar source-region structure leave a discernible imprint on these turbulence properties? Analysis of PSP observations between 0.15 and 0.5 au, spanning half an activity cycle from minimum to maximum, gives an unambiguous answer to the first question. As solar activity increases, we observe pronounced differences in the absolute values and the radial evolution of turbulence energy, cross helicity, residual energy, and energy-dissipation rate, with larger values and slower radial decay of these turbulence properties during high activity; the turbulent correlation scale exhibits relatively faint variation. The mean/large-scale speed and magnetic-field magnitude follow a qualitatively similar trend of enhancement with activity, while proton density is diminished. These results extend previously reported solar-cycle variations in turbulence properties and phenomenological cascade rates at 1 au \citep{zhao2018ApJcr2,gautam2024solar}, while also demonstrating that the patterns observed near Earth are already established much closer to the Sun.
}

\added{One can reasonably expect these trends to be driven by corresponding changes in the global magnetic configuration at the source regions of the wind. Coronal holes, which tend to be located in the Sun's polar regions during sol-min, occur with greater frequency at low heliolatitudes during sol-max, where they can produce fast wind that is sampled by PSP in the ecliptic \citep[see also][]{mccomas2003grl}. At the same time, due to the greater complexity and temporal variability of the source-region distribution during sol-max, one could expect greater mixing of wind streams from diverse source regions; this can potentially dilute the imprint of source structure on the wind observed in the inner heliosphere \citep{schwenn2006solar,Viall2020JGR}.}

\added{While such complicating factors make it challenging to address our second question above, we have made a preliminary attempt in this study at establishing links between in situ observations of turbulence properties and source-region structure. We examine the inferred solar sources of three PSP encounters that sample low (E6), increasing (E14), and high (E22) activity epochs, employing three corresponding simulation runs of the \cite{Usmanov2025ApJ} global model to establish magnetic connectivity. The turbulence energy and cross helicity are radially detrended using empirical fits to account for their secular trends with helioradius. The magnetic footpoints of PSP intervals are sorted into subsets based on these detrended turbulence diagnostics, and these footpoints are overlaid on SDO/AIA 193 \AA\ EUV maps and photospheric magnetograms in order to evaluate source structure.}

\added{The low-activity case suggests that intervals with large turbulence energy and cross helicity tend to be associated with relatively darker regions on the EUV map, and this pattern is much stronger for the high-activity case. These findings suggest that, at least during some epochs, source region structure leaves clear imprints on turbulence observed in the young solar wind. Open, unipolar magnetic topology (associated with dark coronal-hole interiors in EUV) tend to produce wind streams having the strongest fluctuation energies and Alfv\'enicity; in contrast, multipolar magnetic topology (associated with bright active-region environments in EUV) tend to be associated with intervals with smaller values for these turbulence diagnostics.}

\added{However, the E14 case demonstrates that the generality of such patterns cannot be assumed. During this period of increasing solar activity, wind streams of varying turbulence levels are apparently produced by both types of source structure identified above. We have conducted additional magnetic connectivity analysis for several additional PSP encounters (not shown here) using the potential field source surface model with ballistic mapping \citep[e.g.,][]{badman2023JGR}, and these preliminary tests reinforce our stated conclusions: the additional sol-min cases indicate a pattern similar to that observed during E6, but the additional high-activity encounters we analyzed indicated a mix of the contrasting patterns we observed for E14 and E22. A useful future direction would be to incorporate data from PSP's perihelia, which extend as low as 0.046 au, to examine whether the imprints of source structure are stronger in the upper corona and in the sub-Alfv\'enic regime \citep{bandyopadhyay2022ApJ,Badman2026ApJ}.}

\added{In the cases where footpoint EUV intensity was (anti-) correlated with our solar-wind turbulence diagnostics, we did not find a similar correlation with the corresponding footpoint magnetic-field strength. While the EUV images we analyzed are widely used to infer magnetic structure, it will be worthwhile in future work to examine the relationship between solar-wind turbulence and additional quantitative metrics of magnetic field structure, topology, and complexity at the inferred solar sources \citep{Longcope2005LRSP,Shi2025ApJS}. Establishing robust associations could provide a practical means of using readily-computed turbulence parameters to infer the likely source types of in-situ solar wind observations. This would be particularly valuable at smaller helioradii: at 1 au, speed is often a useful, although imperfect, discriminator between wind streams associated with coronal-hole and streamer-belt/closed-loop source regions; closer to the Sun, however, speed is less informative because the fast wind has not yet accelerated to its asymptotic velocity and different source populations may not be well separated in speed \citep[see also][]{Alterman2025ApJ}.}

Another practical application of our results is the improved data-driven specification of mean free paths (MFPs) that characterize diffusion of SEPs and magnetic field lines in the heliosphere \citep{Engelbrecht2022SSR}. Variability in the underlying turbulence properties can, in turn, produce substantial variation in inferred MFPs \citep[e.g.,][]{chhiber2017ApJS230,chhiber2021ApJ_flrw,Lang2024ApJ}. Robust relationships between these properties and solar activity or solar-wind source structure could therefore provide physically motivated constraints on event-specific MFPs prescribed in SEP transport models, particularly when the required turbulence parameters are not directly available from observations \citep{Whitman2023AdSpR}.

\added{To supplement our primary observational analysis, we examined the radial evolution of turbulence energy between the coronal base and the PSP trajectory in the three global model runs that were previously employed to establish magnetic connectivity. For all three solar-activity cases,} the modeled turbulent-energy density evolves radially as the solar wind propagates outward, and reaches a maximum in the sub-Alfv\'enic regime, generally consistent with previous modeling studies \citep{verdini2009ApJ,chhiber2019psp1,Usmanov2025ApJ}. The high-activity case shows a broader and more outward-displaced distribution of locations of maximum turbulence amplitude, closer to the Alfv\'en surface. In the low-activity case the turbulence amplitude peaks deeper in the sub-Alfv\'enic corona.

\added{In summary, our study employs the extensive in situ dataset accumulated by PSP to demonstrate substantial solar-cycle modulation of turbulence in the young solar wind. The observed variability in key turbulence parameters, including fluctuation amplitude and residual energy, has direct implications for scattering of SEPs \citep{Engelbrecht2022SSR}, which pose a significant space weather hazard  particularly during high solar activity \citep{Whitman2023AdSpR}.
Our investigation of the relationship between in-situ PSP observations and inferred solar source regions suggests that source structure can have a significant influence on the variability of turbulence properties in the young solar wind \citep[see also][]{Yardley2024NatAs}. However, we also identify cases in which the turbulence  exhibits no clear imprint of source structure. Together, these findings bear on the long-standing debate on the relative roles of solar origins and local evolution in determining solar wind properties \citep{Viall2020JGR}.
Broader statistical studies, complemented by detailed observational and modeling analyses of individual solar wind streams, will be needed to clarify this complex relationship. The relevant physical mechanisms undoubtedly involve the coupling of large-scale coronal structure with turbulence dynamics \citep{Wang1991ApJ,cranmer2007ApJS,Usmanov2025ApJ,Squire2026arXiv}. Improved theoretical descriptions and numerical models of this coupling are therefore crucial for understanding the origin, evolution, and dissipation of turbulence in the heliosphere.}

\begin{acknowledgments}
This work was supported at the University of Delaware in part by the PSP/IS\(\odot\)IS mission through subcontract SUB0000165 from Princeton, and the PUNCH mission through subcontract N99054DS from NASA/SWRI. This work was also partially supported by the U.S. National Science Foundation, award PHY-2108834, through the NSF/DOE Partnership in Basic Plasma Science and Engineering, by the NASA  Living With a Star (LWS) Science program grant 80NSSC22K1020, and the NASA Heliophysics Supporting Research program grant 80NSSC22K1639. It utilized resources provided at the National Energy Research Scientific Computing Center (NERSC).
R.C. acknowledges valuable discussions with Barbara J. Thompson and Teodora Mih\v ailescu. R.C. and W.H.M. acnowledge the support of the International Space Science Institute (ISSI, Bern), through International Team project No. 24–608—``Energetic Particle Transport in Space Plasma Turbulence''.

\end{acknowledgments}

\appendix

\section{PSP data processing and interval selection}\label{app:psp-data}

We analyze Parker Solar Probe (PSP) observations from orbits 1-25, covering 2018 October 31 00:00:00 to 2025 November 1 00:00:00 UTC. 
Proton plasma moments are taken from the Solar Probe Cup (SPC) of the SWEAP suite \citep{kasper2016SSR,case2020ApJS}, and magnetic-field measurements are taken from the fluxgate magnetometer (MAG) of the FIELDS suite \citep{bale2016SSR}. 
We use PSP SWEAP/SPC Level-3 ion moments and PSP FIELDS/MAG Level-2 RTN magnetic-field data. 
All plasma velocities and magnetic-field vectors are expressed in RTN coordinates \citep{franz2002pss}.

SPC ion moments are screened using fill values, physical range checks, and the mission-provided quality information \citep{SWEAP2020UserGuide}. 
Samples failing the selected quality criteria are excluded by setting the proton density and all three velocity components to \(NaN\). 
Magnetic-field samples are screened using fill values and physical range checks.

To compute quantities requiring simultaneous plasma and magnetic-field measurements, the MAG time series is linearly interpolated onto the SPC time stamps after invalid MAG samples are masked. 
An interpolated magnetic-field value is retained only when the corresponding SPC time lies within the valid MAG time-range and the nearest valid MAG sample is separated by no more than 5 s; otherwise, the interpolated value is discarded.

The cleaned data are divided into non-overlapping 2-hr intervals. 
We retain intervals for which at least 60\% of the SPC samples contain valid proton density and all three components of the proton bulk velocity. 
For turbulence quantities, we further require at least 80\% of SPC time stamps in the interval to have simultaneous valid density, velocity, and gap-aware interpolated magnetic field.
For each 2-hr interval, we use the heliocentric distance at the temporal midpoint for radial binning and field-line tracing.
Unless stated otherwise, the statistical analysis uses intervals with heliocentric distance $0.15 \leq r < 0.50$ AU.
The lower bound avoids the near-Sun sub-Alfv\'enic regime, where the Taylor hypothesis used to estimate $\lambda$ becomes increasingly uncertain \citep{chhiber2019psp2}. 
Another reason to maintain this lower bound was to the avoid the requirement of switching from SPC to SPAN data at \(r \lesssim 20~R_\odot\), which would introduce complications related to incomplete sampling of ion distributions within the SPAN field-of-view \citep{kasper2016SSR,Badman2025ApJ}.

The solar-wind speed $V$ and magnetic-field strength $B$ plotted in Fig. \ref{fig:hist2d_var} are the magnitudes of the 2-hr mean RTN vectors. 
The proton density $n_{\mathrm{p}}$
is the average of an interval over valid SPC samples after quality screening.

\section{Turbulence Diagnostics}\label{app:turbulence-diagnostics}

We employ standard approaches to compute turbulence parameters \citep{pope2000book,bruno2013LRSP,chhiber2021ApJ_psp}. Within each retained 2-hr interval, velocity and magnetic fluctuation vectors are defined relative to the interval mean vectors: e.g., \(\delta \mathbf{v}=\mathbf{V}-\mathbf{V}_0\), where \(\mathbf{V}_0\) is the interval mean. 
Turbulence averages are evaluated only over samples for which the proton density, all three velocity components, and the interpolated magnetic-field vectors are simultaneously valid. 
We define the Alfv\'en speed as $V_A = B/\sqrt{\mu_0 \rho_{\mathrm{p}}}$ and express magnetic-field fluctuations in the corresponding velocity units as $\delta \mathbf{b}_A = \frac{\delta \mathbf{B}}{\sqrt{\mu_0 \rho_{\mathrm{p}}}}$, where $\rho_{\mathrm{p}}$ is the interval-mean proton mass density.
The Els\"asser fluctuations are $\delta \mathbf{z}^{\pm} = \delta \mathbf{v} \pm \delta \mathbf{b}_{\rm A}$.
The total fluctuation variance in velocity units, used here to measure the turbulence energy density, is defined as:
\begin{equation}
    Z^2
    =
    \frac{1}{2}
    \left\langle
    |\delta \mathbf{z}^{+}|^2
    +
    |\delta \mathbf{z}^{-}|^2
    \right\rangle
    =
    \left\langle
    |\delta \mathbf{v}|^2
    +
    |\delta \mathbf{b}_{\rm A}|^2
    \right\rangle ,
\end{equation}
where $\langle * \rangle$ denotes an average over valid simultaneous plasma-field samples in the 2-hr interval, and $Z=\sqrt{Z^2}$.
This is (twice) the total kinetic-plus-magnetic fluctuation energy per unit mass.

The normalized cross helicity is \(\sigma_c^\text{raw}=2\langle\delta\mathbf{v}\cdot\delta\mathbf{b}_\text{A}\rangle Z^{-2}\), a measure of the degree of correlation of the \(V\) and \(B\) fluctuations. Equivalently,
\begin{equation}
    \sigma_c^{\rm{raw}}
    =
    \frac{
    \left\langle |\delta \mathbf{z}^{+}|^2
    -
    |\delta \mathbf{z}^{-}|^2
    \right\rangle
    }{
    \left\langle |\delta \mathbf{z}^{+}|^2
    +
    |\delta \mathbf{z}^{-}|^2
    \right\rangle
    }.
\end{equation}
For statistical analyses across intervales, we use a magnetic-polarity-rectified helicity, denoted $\sigma_c^\text{rect}$. 
For our Els\"asser-variable convention, the unrectified value is multiplied by $-\mathrm{sgn}(\langle B_R\rangle)$, removing the magnetic-sector-dependent sign reversal so that positive $\sigma_c$ consistently denotes dominance of anti-sunward propagating Alfv\'enic fluctuations \citep[e.g.,][]{roberts1987JGRb}.
In this paper, $\sigma_c$ denotes $\sigma_c^{\rm{rect}}$ unless
otherwise stated.
The normalized residual energy is
\begin{equation}
    \sigma_D
    =
    \frac{
    \left\langle |\delta \mathbf{v}|^2
    -
    |\delta \mathbf{b}_{\rm A}|^2
    \right\rangle
    }{
    \left\langle |\delta \mathbf{v}|^2
    +
    |\delta \mathbf{b}_{\rm A}|^2
    \right\rangle
    },
\end{equation}
which represents the energy partition between velocity and magnetic fluctuations, and is zero for an Alfv\'en wave.
Positive values indicate velocity-dominated fluctuations, whereas negative values indicate magnetic-fluctuation dominance \citep{bruno2013LRSP}.

The correlation length $\lambda$ is estimated independently in each 2-hr interval from the $1/e$ decay time of the magnetic-field autocorrelation function (ACF), following standard solar-wind turbulence analysis \citep{matthaeus1982JGR,Ruiz2014SoPh,roy2021ApJl}, and it serves as a representative energy-containing similarity scale.
We first require the MAG RTN components to be at least 90\% valid in the 2-hr interval. 
The magnetic-field time series is then interpolated onto a 1-s uniform grid across short data gaps only (less than 5 s),  and the ACF of the mean-subtracted magnetic-field fluctuations is computed as
\begin{equation}
    C_B(\tau)
    =
    \frac{
    \sum_i
    \left\langle
    \delta B_i(t)\,\delta B_i(t+\tau)
    \right\rangle
    }{
    \sum_i
    \left\langle
    \delta B_i^2(t)
    \right\rangle
    },
\end{equation}
where $i$ denotes the RTN components. 
The correlation time $(t_{\rm corr})$ is defined as the first lag at which $C_B(\tau)$ falls to $1/e$, using linear interpolation between adjacent lag samples. 
The correlation length is then obtained as $\lambda=Vt_{\rm corr}$ using Taylor's frozen-in-flow hypothesis \citep{taylor1938ProcRSL}, where $V$ is the 2-hr mean solar-wind speed.
This speed $V$ is used for the correlation-length estimate only when at least 50\% of velocity vectors are valid in the interval. Note that the measured $\lambda$ is evaluated along the plasma-advection direction in the spacecraft frame; since the angle between the sampling direction and the local mean magnetic field varies throughout the PSP orbit, this estimate cannot be identified uniquely with correlations either parallel or perpendicular to the field \citep{dasso2005ApJ,Cuesta2022ApJL}. The solar-activity variation of turbulence anisotropy is deferred to future study.

We next estimate the energy transfer rate using von K\'arm\'an decay phenomenology, modified for MHD with finite cross helicity \citep{zhou2004RMP}.
Let $Z_\pm^2 \equiv \langle |\delta \mathbf{z}^{\pm}|^2\rangle$, then $Z_\pm=\sqrt{Z_\pm^2}$ denote the Els\"asser variables.
Their general phenomenological decay is written as
\begin{equation}
    -\frac{d Z_\pm^2}{dt}
    =
    \alpha_\pm
    \frac{Z_\pm^2 Z_\mp}{\lambda_\pm},
\end{equation}
where $\lambda_\pm$ are the corresponding energy-containing similarity scales and $\alpha_\pm$ are dimensionless decay coefficients \citep{hossain1995PhFl, wan2012JFM697}.
In general, $\lambda_+$ and $\lambda_-$ need not be equal.
Following the single-scale closure adopted in solar-wind turbulence-transport models, we approximate $\lambda_+\simeq\lambda_-\simeq\lambda$ and $\alpha_+\simeq\alpha_-$, using the magnetic-field correlation length estimated above as a representative energy-containing scale \citep{breech2008turbulence,chhiber2021ApJ_psp}.

Using $Z_\pm^2=(1\pm\sigma_c)Z^2$ under this closure gives the standard cross-helicity correction factor \citep{matthaeus2004grl}
\begin{equation}
    f(\sigma_c)
    =
    \frac{1}{2}
    \sqrt{1-\sigma_c^2}
    \left(
    \sqrt{1+\sigma_c}
    +
    \sqrt{1-\sigma_c}
    \right).
\end{equation}
The cross-helicity-modified von K\'arm\'an energy-transfer rate is then
\begin{equation}
    \varepsilon_{\rm VK}
    =
    \frac{f(\sigma_c) Z^3}{\lambda}.
\end{equation}
The factor $Z^3/\lambda$ provides a phenomenological estimate of the turbulent energy transfer rate at the energy-containing scale, whereas $f(\sigma_c)$ accounts for the weakening of nonlinear interactions when one Els\"asser population dominates \citep{podesta2011energy,Chhiber2026ApJ}. 

In the above, we have neglected the dimensionless decay coefficient and the fixed numerical factor associated with the normalization of the specific fluctuation energy as $Z^2/2$.
Since our study focuses on relative solar-cycle variations rather than an absolute value of the energy transfer rate, we do not assign a numerical value to this coefficient, which would remain constant in the classical turbulence description \citep{karman1938prsl,hossain1995PhFl}.
Note that the precise value of the coefficient is a topic of active turbulence research; it may not be strictly universal and can depend on the mean magnetic-field strength, cross and magnetic helicities, forcing, and flow compressibility \citep{linkmann2015nonuniversality, linkmann2017reynolds, li2024non}.
Nevertheless, solar-wind turbulence modeling efforts have consistently employed a constant coefficient \citep{zank1996evolution,vanderholst2014ApJ,usmanov2018,zank2018ApJ}, and accordingly, our analysis assumes that the coefficient varies little with the solar activity.

The von K\'arm\'an formulation assumes locally-homogeneous turbulence and approximate self-preservation of the normalized second- and third-order correlations, with the energy-containing scale well separated from both the dissipation scale and the scale of large-scale system variation \citep{hossain1995PhFl,wan2012JFM697}.
It has been successfully applied to estimate the turbulent energy-transfer rate in the solar wind \citep{bandyopadhyay2020ApJS_cascade,Wu2022ApJ}. We employ it here as a convenient alternative to the (more exact) third-order law estimate of the inertial-range cascade rate \citep{politano1998PRE}.
\section{Relating sunspot number to PSP radial distributions}\label{app:ssn-binning}
\begin{figure}
    \centering
    \includegraphics[width=0.5\linewidth]{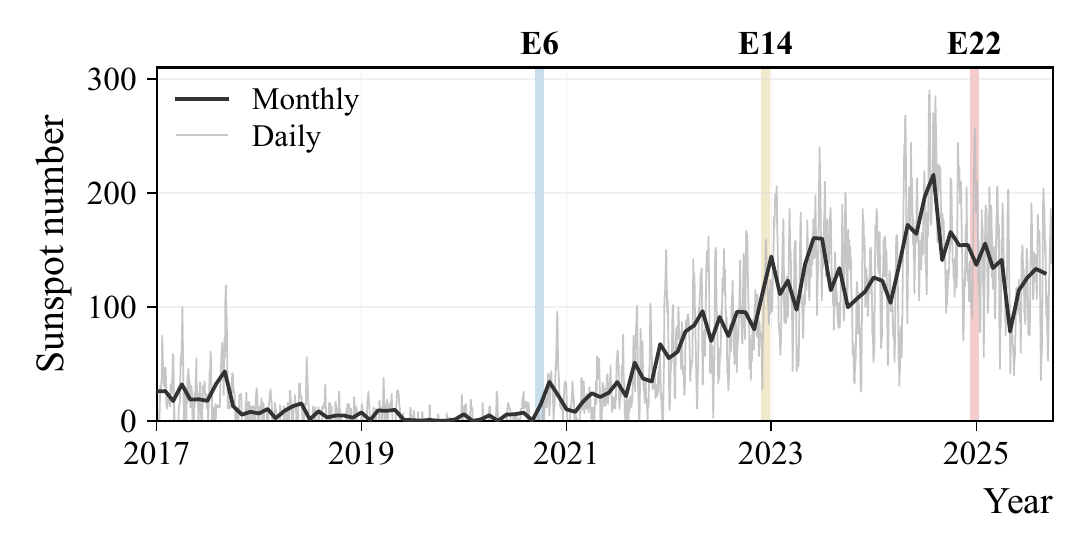}
    \caption{Sunspot number evolution across approximately half a solar activity cycle. The three PSP encounters analyzed in this paper are marked.}
    \label{fig:sunspotn}
\end{figure}

Each 2-hr PSP interval is matched, without temporal averaging or interpolation, to the tabulated daily SSN for the UTC date from the World Data Center SILSO  \citep{SILSO_Sunspot_Number}. \added{For convenient reference, Fig.~\ref{fig:sunspotn} illustrates the evolution of SSN from sol-min to sol-max.}
Intervals whose midpoint day does not have a valid daily SSN are excluded from the SSN-binned distribution of Fig. \ref{fig:hist2d_var}.
The $(r,y)$ plane is binned for each plotted quantity $y$. 
Radial bins have width $\Delta r=0.05$ AU and cover $0.15\leq r<0.50$ AU. 
The vertical coordinate is divided into 20 variable-specific bins over the displayed range.
Logarithmic spacing is used for $Z^2$, $n_{\rm p}$, $\lambda$, and $\varepsilon_{\rm VK}$, whereas linear spacing is used for $V$, $B$, $\sigma_c$, and $\sigma_D$.
Bins containing fewer than five 2-hr intervals are not displayed.

For each displayed $(r,y)$ cell, the color represents the mean of the daily SSN values assigned to the 2-hr intervals in a cell.
The overlaid low and high activity radial trends are computed from the displayed two-dimensional bin values.
Displayed bins with mean $\mathrm{SSN}<75$ define the low-activity subset, and those with mean $\mathrm{SSN}>80$ define the high-activity subset. 
In each radial column, the trend value is the count-weighted mean of the bin-mean $y$ values, where the weight of each bin is the number of 2-hr intervals that it contains.
The vertical bars show the corresponding count-weighted 25-75\% range of the displayed bin means.

To validate the robustness of the results, we repeated the analysis using the median, rather than the mean, of the interval-assigned daily SSN values within each $(r,y)$ cell.
We also performed the same analysis directly using the individual 2-hr intervals (instead of the bins), and computed the low and high activity radial trends from the interval-level distributions.
In both cases, the comparisons between low and high activity are qualitatively unchanged.
\section{Global MHD model, magnetic connectivity, and solar source regions}
\label{app:mhd-connectivity}
\begin{figure}
    \centering
    \includegraphics[width=.32\linewidth]{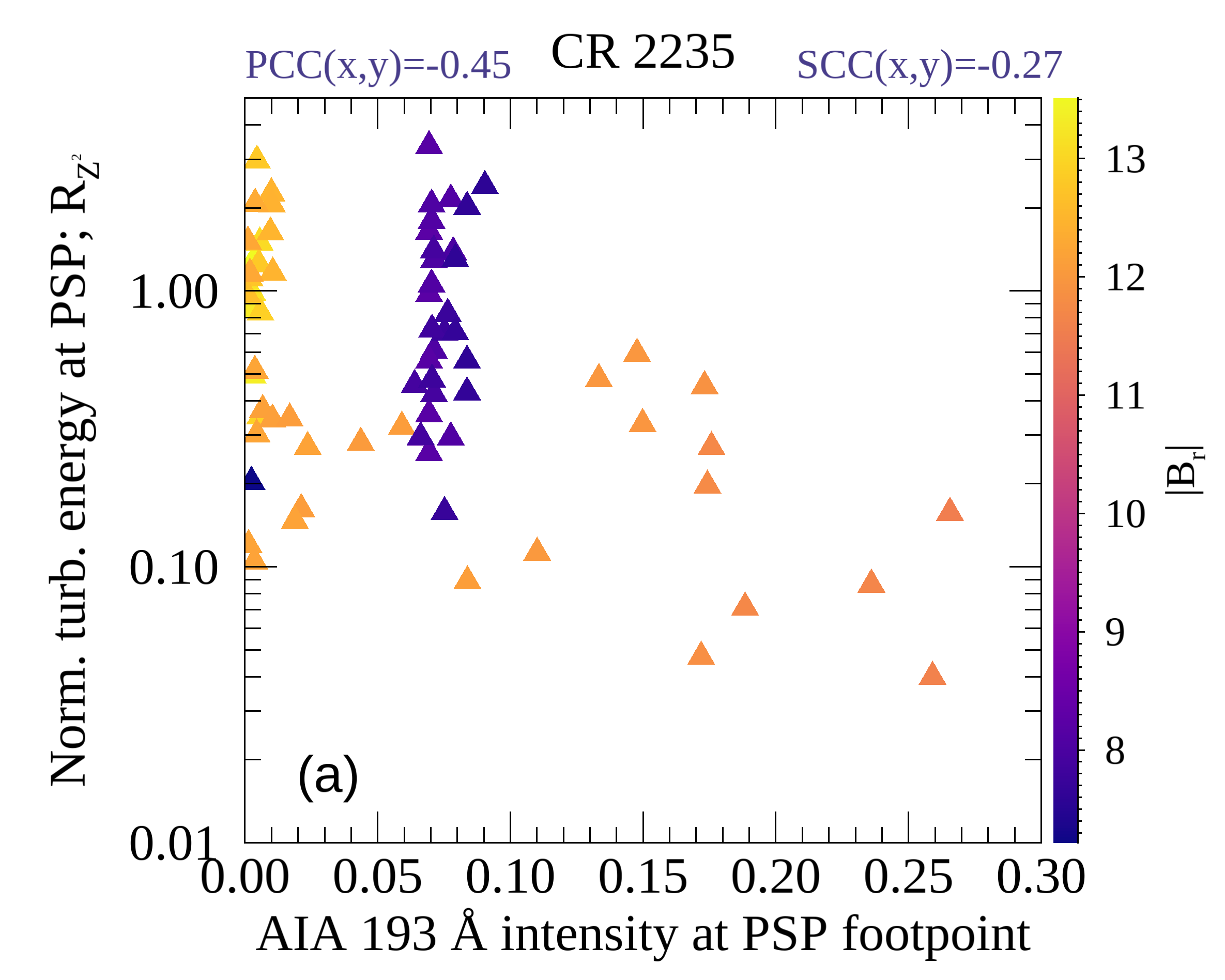}
    \includegraphics[width=.32\linewidth]{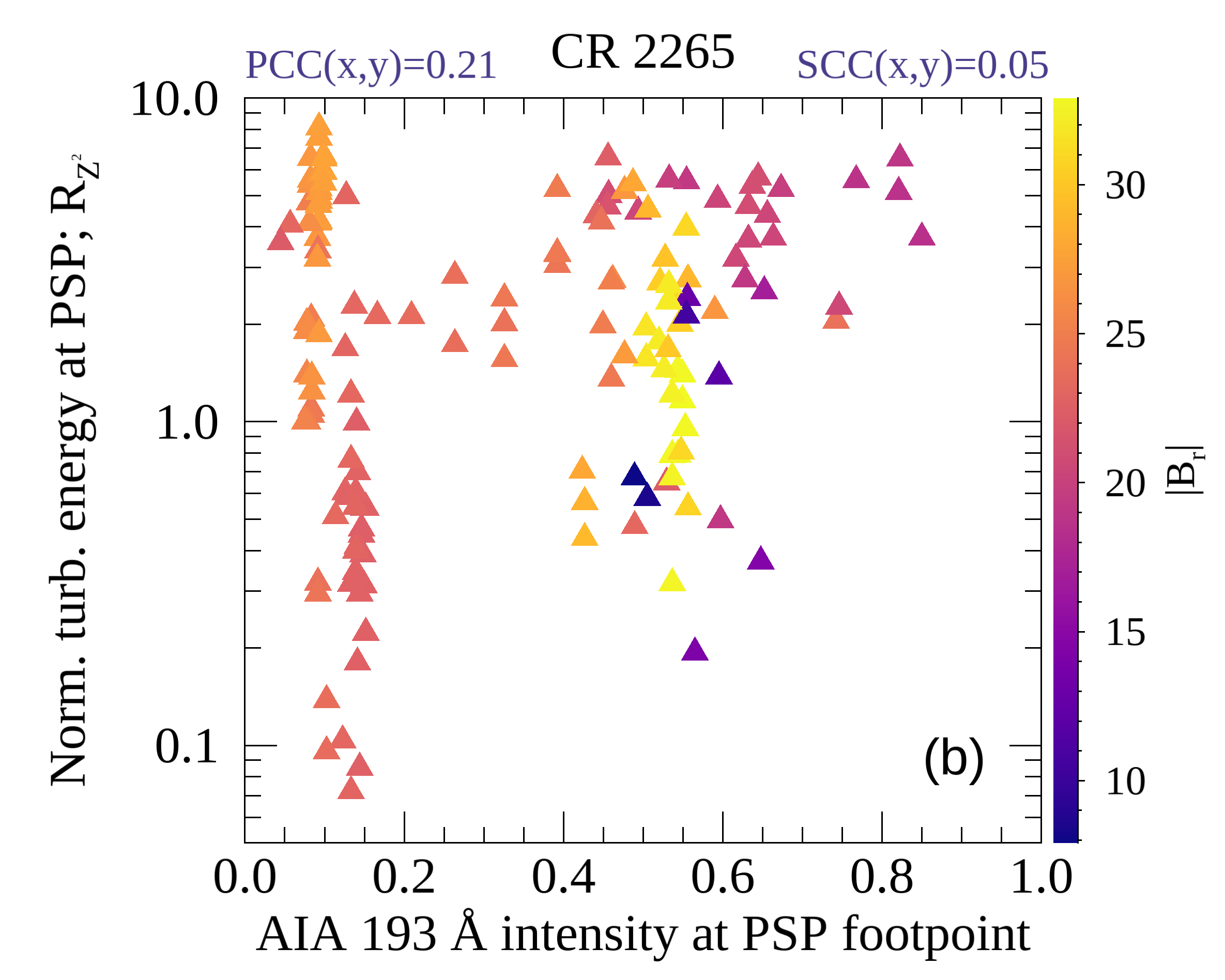}
    \includegraphics[width=.32\linewidth]{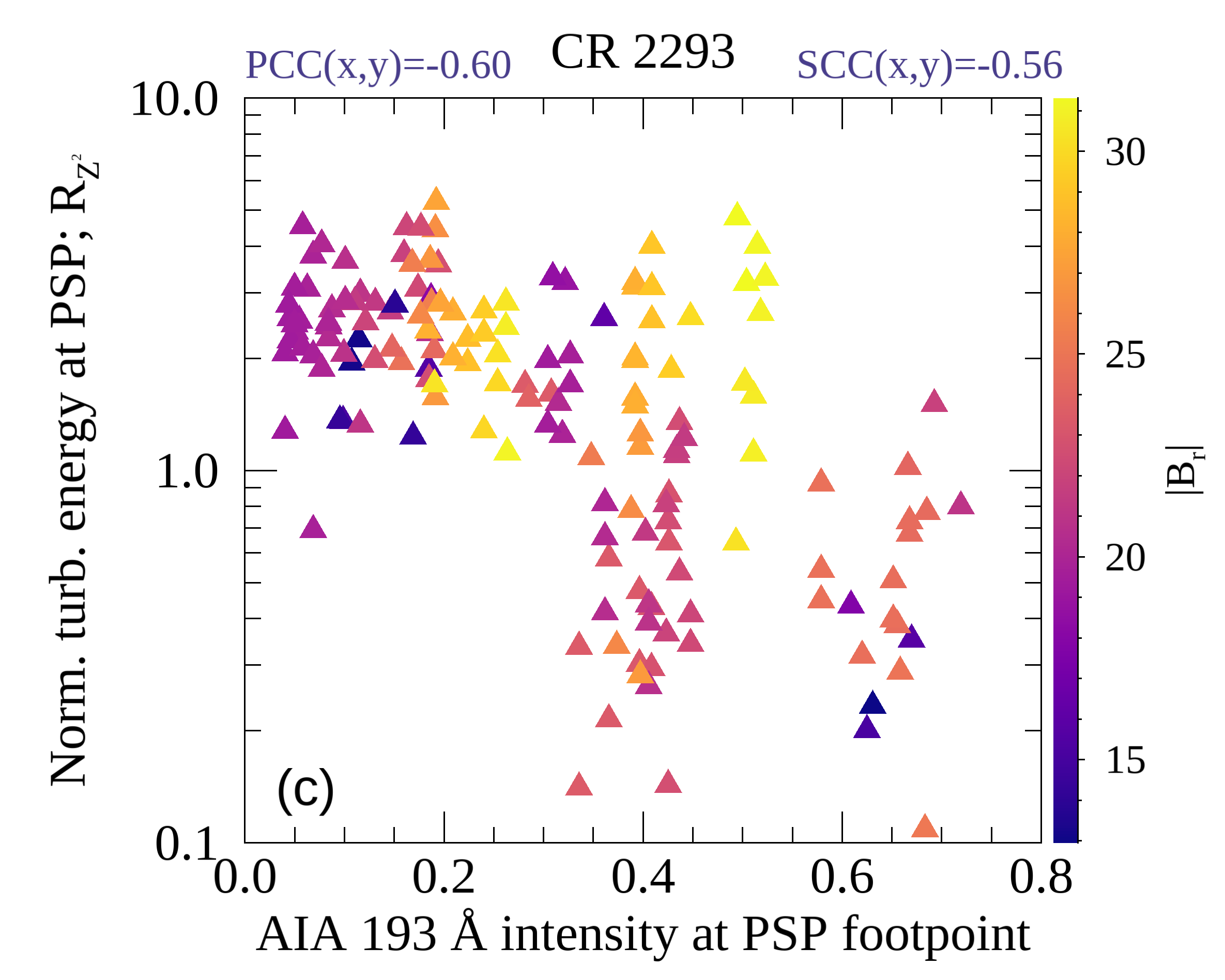}
    \includegraphics[width=.32\linewidth]{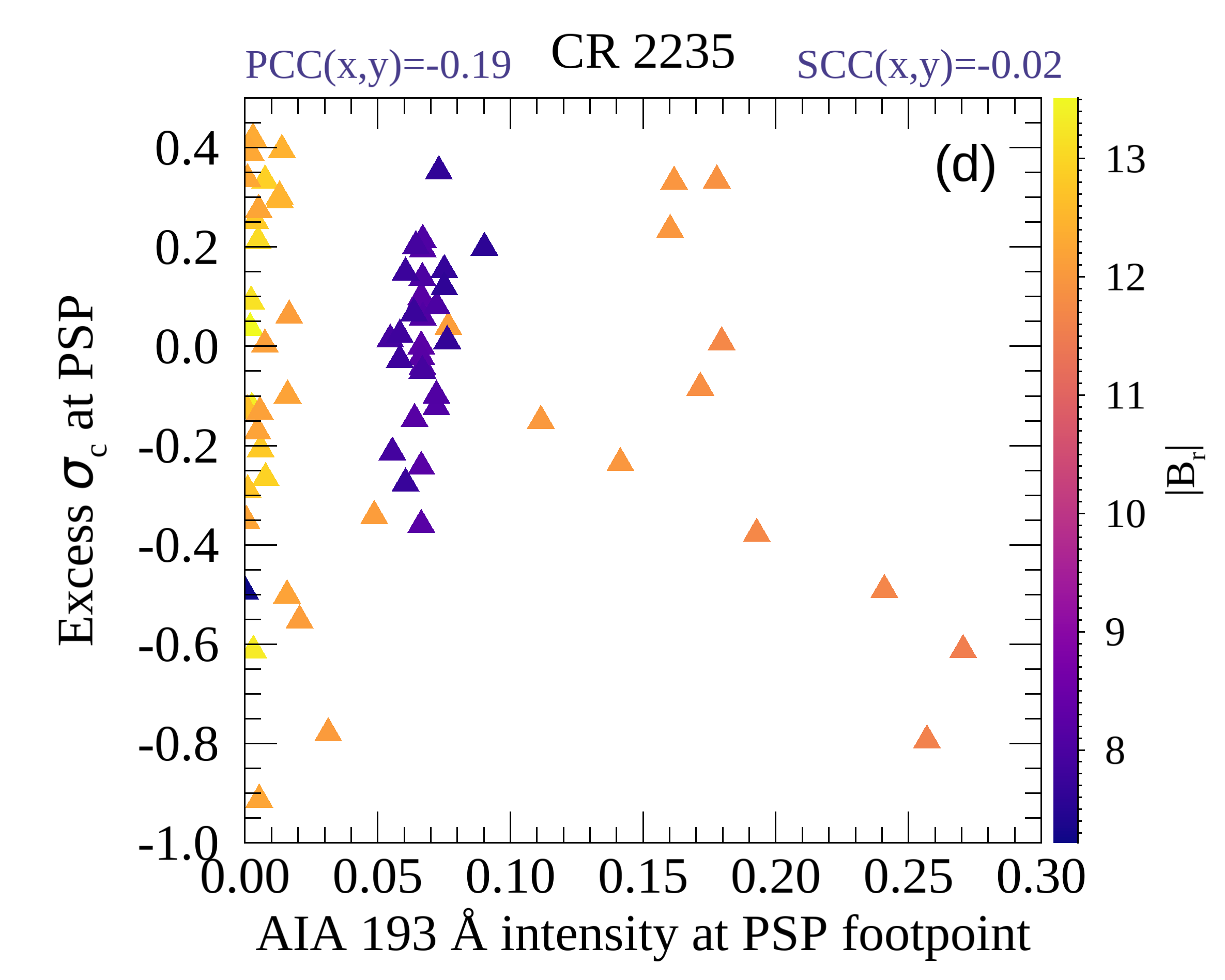}
    \includegraphics[width=.32\linewidth]{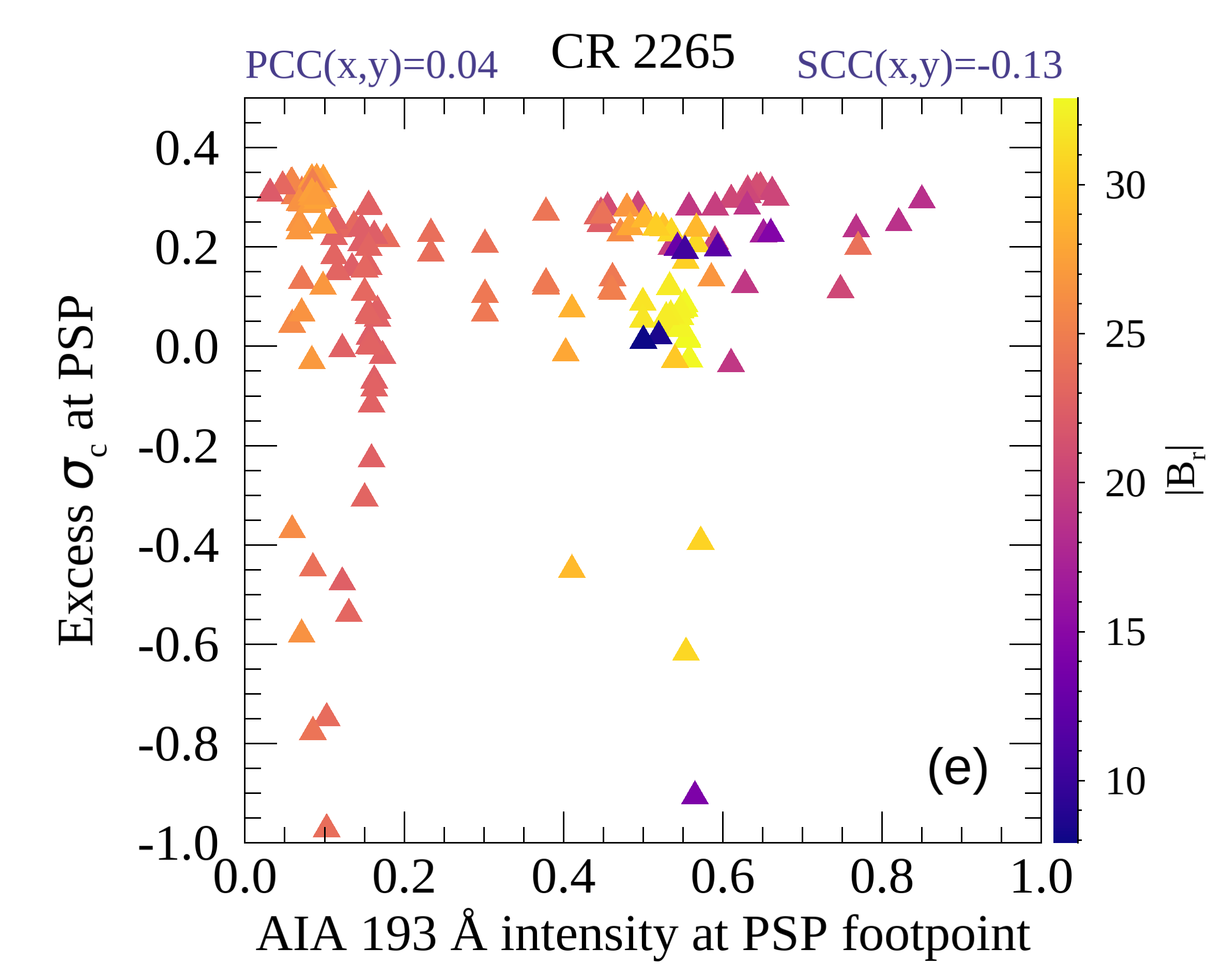}
    \includegraphics[width=.32\linewidth]{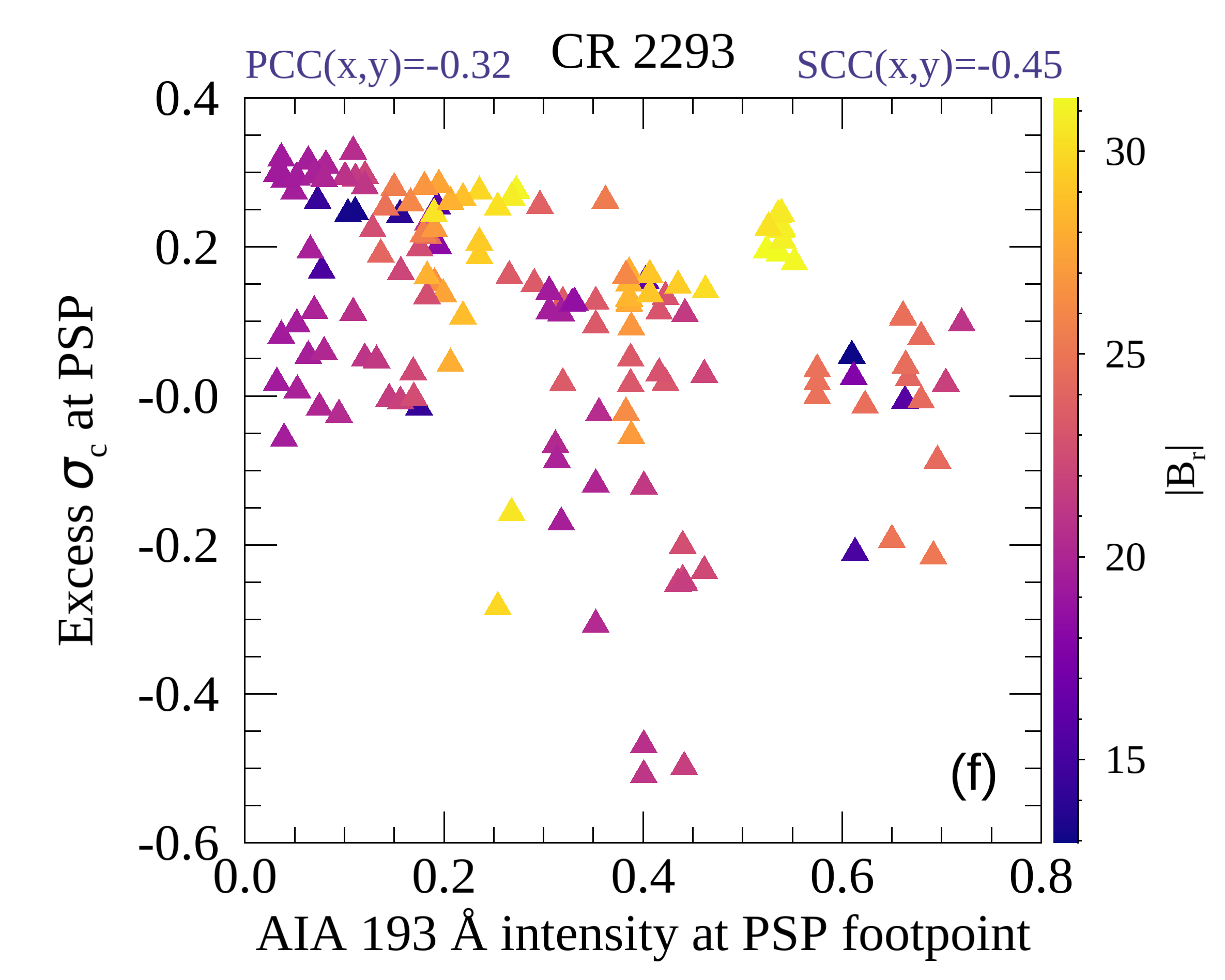}
    \caption{PSP observations of (top) \(R_{Z^2}\) and (bottom) \(\Delta\sigma_c\) scatter-plotted against the AIA 193-\AA~intensity at the magnetic footpoints of the respective PSP intervals. Left, middle, and right columns are from E6 (CR 2235), E14 (CR 2265), and E22 (CR 2293), respectively. The symbols are colored by the radial magnetic field strength (\(|B_r|\)) at the respective footpoints, with its values shown in the respective colorbars. The Pearson [PCC\((x,y)\)] and Spearman [SCC\((x,y)\)] correlation coefficients of the abscissae (\(x\)) and the ordinates (\(y\)) are annotated at the top of each panel; here \(y\) denotes \(\log \RZ\) and \(\Delsig\) in the top and bottom panels, respectively, while \(x\) denotes AIA intensity.
}
    \label{fig:AIA_intensity}
\end{figure}
4HJ-HBL
We employ the Usmanov et al. global 3D MHD model of the corona and inner heliosphere \citep{Usmanov2025ApJ}. The model couples Reynolds-averaged mean-field equations for bulk velocity, density, magnetic field, and temperature to transport equations for statistical turbulence parameters, including \(Z^2,~\lambda, \sigma_c,\) and \(\sigma_D\). This coupling is able to effectively model the heating and acceleration of the solar wind by turbulence \citep{matthaeus2011SSR}, an approach widely employed in global heliospheric modeling \citep{gombosi2018LRSP}. The computational domain is divided into two sub-regions. The inner (coronal) sub-region extends from the coronal base at \(1\ R_\odot\) to \(30\ R_\odot\); the outer (solar wind) region extends from \(30\ R_\odot\) to 5 au. Steady-state solutions are obtained in the inner region by time relaxation, starting from an initial state composed of either a Parker-type flow in a dipole magnetic field \citep{usmanov2003tilted} or derived from synoptic solar magnetograms for different solar rotations \citep{usmanov2014three,usmanov2018,Usmanov2025ApJ}. 
GONG/ADAPT synoptic maps \citep{arge2010AIPC} are used in the present work. Boundary conditions for the outer region are extracted from the inner region solution, and steady state solutions are again obtained by time relaxation. The resolution of standard runs is \(702\times 120\times 240\) along \(r\times \theta\times \phi\) coordinates (between \(1~R_\odot\) and 5 au). The computational grid has logarithmic spacing along the radial (\(r\)) direction with the grid spacing becoming larger as \(r\) increases. The latitudinal (\(\theta\)) and longitudinal (\(\phi\)) grids have equidistant spacing with a resolution of 1.5\degree~each. The grid spacing is typically about a few times larger than the correlation scale of turbulence \citep{Ruiz2014SoPh,Cuesta2022ApJL}. 

\added{Note that our use of the photospheric synoptic map at the coronal base is intended as an approximate observational constraint on the magnetic field, following a widely used approach in global coronal modeling \citep[e.g.,][]{usmanov2014three,Shiota2017ApJ837,MacNeice2018SpWea,Reiss2019ApJS,reville2020ApJS}. Such models do not resolve the narrow region between the photosphere and the coronal base, which includes the chromosphere and transition region (TR), across which the density and temperature undergo steep gradients and Alfv\'en waves can experience substantial reflection \citep[e.g.,][]{cranmer2005apjs}. While approximate numerical approaches for resolving this region have been developed \citep[][and references within]{gombosi2018LRSP}, there is evidence that the large-scale magnetic structure remains substantially similar in simulations with or without the TR \citep{Lionello2001ApJ,Downs2010ApJ}; our main purpose for the \cite{Usmanov2025ApJ} model in the present work is to obtain precisely this large-scale magnetic field distribution. Note that all other inner boundary conditions in the model are specified consistently for the coronal base.} For further details on our model, including initial and boundary conditions, we refer the reader to \cite{Usmanov2025ApJ}, emphasizing here that it is mature, well-tested, and has been validated by comparison with a variety of heliospheric observations \citep[e.g.,][]{usmanov2018,bandyopadhyay2020ApJS_cascade,chhiber2021ApJ_psp,Usmanov2025ApJ}.

\added{In the present study, we employ three simulation runs for case studies of low, increasing, and high solar activity epochs. The only difference between these runs lies in the synoptic map of the magnetic field used as the boundary condition at the coronal base: CRs 2235, 2265, and 2293, corresponding to PSP E6, E14, and E22, respectively.}

\added{To establish magnetic connectivity to the coronal base} for each 2-hr interval, first the PSP trajectory location is evaluated at the interval midpoint by linear interpolation.
The model magnetic field line passing through this position is numerically integrated towards the Sun. The field-line tracing software employed here uses the Runge-Kutta-Merson method \citep{lance1960numerical} and has been adapted from the GEOPACK package developed by \cite{Tsyganenko_GEOPACK}.

The surface footpoints are overlaid on 193 \AA\ extreme-ultraviolet (EUV) images of the solar corona (Figs. \ref{fig:2235_aia_ftpt}, \ref{fig:2265_aia_ftpt}, \ref{fig:2293_aia_ftpt}). We employed synoptic maps corresponding to CR 2235, \added{2265,} and 2293, produced by the Atmospheric Imaging Assembly (AIA) instrument aboard NASA's Solar Dynamics Observatory \cite[SDO;][]{Lemen2012SoPh}. \added{We obtained these maps from NJIT's IDSEAR catalog \citep{Illarionov2020ApJ}.}
The maps have a resolution of \(720 \times360\), uniformly spaced in solar longitude and latitude. These images carry information about coronal structures such as active regions, coronal holes, and large-scale magnetic-field-associated features \citep{DelZanna2018LRSP_Solar_UV_Xray}.

Fig. \ref{fig:AIA_intensity} shows PSP observations of the normalized turbulence-energy density and excess \(\sigma_c\), scatter-plotted against the AIA 193 \AA\ intensity at the respective magnetic footpoints, in separate panels for each analyzed PSP encounter. \(Z^2\) and \(\sigma_c\) have been radially detrended as described in App. \ref{app:relative-turbulence-selection} and in the main text. For this plot, the AIA 193-\AA~intensity of a footpoint is computed as a 9-pixel 2D-boxcar average centered on the respective footpoint; this average provides a sense of the very-local EUV environment of the low-coronal sources of PSP intervals. \added{In the sol-max case (right panels) a clear decreasing trend is apparent in both \(R_{Z^2}\) and \(\Delta\sigma_c\) as footpoint EUV intensity increases. The trend in the sol-min case (left panels) is weaker, and is apparently completely absent in the E14 case (middle panels).  Computed correlation coefficients are stated in the caption. The symbols are colored by radial magnetic field strength (\(|B_r|\)) at the respective footpoints; this does not appear to be correlated with either \(\RZ\) or \(\Delsig\) in any case.}\footnote{\added{The Pearson and Spearman correlation coefficients with respect to footpoint \(|B_r|\) are noted here as (PCC, SCC), with the first and second pairs of each CR corresponding to \(\RZ\) and \(\Delsig\), respectively. CR 2235: \((-.18,-.05),~(.17, .04)\);  CR 2265: \((.10,.10),~(.00, -.06)\); CR 2293: \((.12,.04),~(.09,.12)\).}}

\begin{figure}
    \centering
    \includegraphics[width=0.5\linewidth]{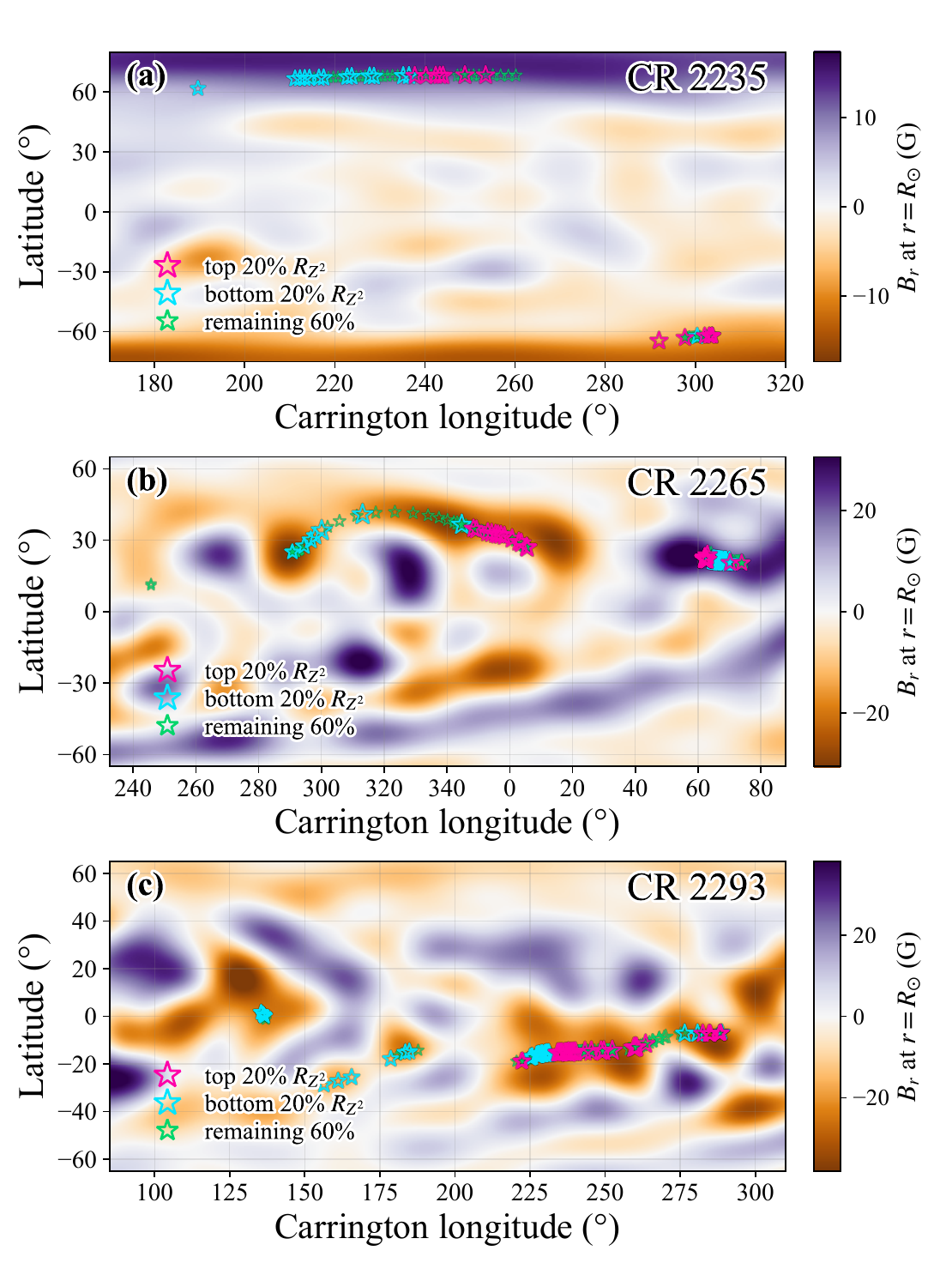}
    \caption{Photospheric sources of PSP observations during E6, E14, and E22, corresponding to CR 2235 (top), CR 2265 (middle), and CR 2293 (bottom).
    The background shows the radial magnetic field $B_R$ at the model inner boundary $r=R_{\odot}$. 
    Magenta, green, and cyan star symbols denote footpoints of solar wind streams with high, middle, and low turbulence amplitude, as described in the caption of Fig.~ \ref{fig:2235_aia_ftpt}.
}
    \label{fig:trace_Br}
\end{figure}

To further evaluate the connections of the PSP observations to the source magnetic configuration \added{(see discussion and caveats above)}, the traced footpoints are overlaid on the radial magnetic field $B_r$ at the model inner boundary $r=R_{\odot}$ in Fig.~\ref{fig:trace_Br}. These figures have already been discussed in the Results section (Sec. \ref{subsec:corona_source_region}).

As an independent robustness check, we repeated the footpoint connectivity analysis using a potential-field source-surface (PFSS) extrapolation with a source surface at $2.5~R_{\odot}$ \citep{schatten1969model,altschuler1969magnetic}. 
The PFSS solution recovered the same broad sol-min and sol-max source region patterns, although a small number of individual footpoints differ from the global-MHD mapping.

\section{Radial detrending and definition of turbulence and Alfv\'enicity subsets}
\label{app:relative-turbulence-selection}

The PSP observations show systematic radial variation in both the turbulence energy-density $Z^2$ and the sector-rectified cross helicity $\sigma_c$ \citep{chen2020ApJS,chhiber2021ApJ_psp}. 
To compare intervals from different heliocentric distances, we removed the leading-order radial dependence of each quantity separately for the low and high solar-activity intervals.

For $Z^2$, all valid PSP intervals with $0.15\leq r < 0.50$ AU were fitted using
\begin{equation}
    Z^2_{\rm fit}(r)
    =
    A\left(\frac{r}{r_0}\right)^{-\alpha},
\end{equation}
where $r_0$ is the geometric-mean radius of all the intervals entering each fit.
We define a normalized ratio for each 
turbulent energy-density measurement:
\begin{equation}
    R_{Z^2}
    =
    \frac{Z^2}{Z^2_{\rm fit}(r)}.
\end{equation}
This is effectively a detrended quantity, whose values above or below unity correspond to $Z^2$ larger or smaller than the fitted value at the same heliocentric distance.

As the cross helicity is bounded, we fit its radial variation empirically as
\begin{equation}
    \sigma_{c,\mathrm{fit}}(r)
    =
    c_0+c_1\ln\left(\frac{r}{r_0}\right),
\end{equation}
using the same solar-min and solar-max time ranges and the same radial interval as for the $Z^2$ fits, where $r_0$ is again the geometric-mean radius of the fitted sample.
The fitted values are restricted to the physically allowed interval $[-1,1]$. 
The relative Alfv\'enicity of each interval is then quantified by the residual
\begin{equation}
    \Delta\sigma_c
    =
    \sigma_c-\sigma_{c,\mathrm{fit}}(r).
\end{equation}
Positive $\Delta\sigma_c$ denotes an interval that is more Alfv\'enic than the fitted radial trend, whereas negative $\Delta\sigma_c$ denotes an interval that is less Alfv\'enic than expected at the same radius.

\added{The radial fits are obtained separately from the PSP observations for low, medium, and high activity epochs -- 2018 October 31 to 2021 April 30, 2022 January 1 to 2024 January 1, and 2023 June 23 to 2025 November 1, respectively. We thus obtain three sets of fitting parameters:
\begin{itemize}
    \item $(A, \alpha, r_0)$, for $Z^2$, are $(1575.047, 2.01516, 0.28532)$ for low; $(4050.891, 1.63951, 0.32187)$ for intermediate; $(5000.417, 0.86749, 0.31722)$ for high activity. \(A\) and $r_0$ are in units of $\mathrm{km}^2 \cdot \mathrm{s}^{-2}$ and $\mathrm{AU}$, respectively.
    \item $(c_0, c_1, r_0)$, for $\sigma_c$, are $(0.509326, -0.442062, 0.28532)$ for low; $(0.703164, -0.328952, 0.32187)$ for intermediate; and $(0.718783, -0.242247, 0.31722)$ for high activity. Here \(r_0\) is in units of au.
\end{itemize}
These fits are plotted as a function of time in Figs. \ref{fig:timeser1}, \ref{fig:timeser2}, \ref{fig:timeser3}, and as radial trends in Fig. \ref{fig:Z2_radial_dependence}.}
%




\end{document}